\documentclass[fleqn,usenatbib]{mnras}

\usepackage{newtxtext,newtxmath}

\usepackage[T1]{fontenc}
\usepackage{ae,aecompl}

\usepackage{graphicx}	
\usepackage{amsmath}	
\usepackage{color}
\usepackage{relsize}

\newcommand{\deriv}{\mathrm{d}}
\newcommand{\first}{^{(1)}}
\newcommand{\second}{^{(2)}}

\newcommand{\fourier}{\mathrm{FT}}

\input{journalabbvr.texinc}

\title[Approximate methods with modified gravity]{Fast numerical method to
  generate halo catalogs in modified gravity (part I): second-order Lagrangian
  Perturbation Theory}

\author[C. Moretti et al]{
C. Moretti $^{1,2,3}$ \thanks{E-mail: chiara.moretti@inaf.it}, 
S. Mozzon $^{1,4}$,
P. Monaco $^{1,2,3,5}$,
E. Munari $^{2}$,
M. Baldi $^{6,7,8}$ \\
$^{1}$Dipartimento di Fisica dell'Universit\'a di Trieste, Sezione di
Astronomia, via Tiepolo 11, I-34143 Trieste, Italy\\ $^{2}$INAF --
Osservatorio Astronomico di Trieste, Via Tiepolo 11, I-34143 - Trieste,
Italy\\ $^{3}$IFPU -- Institute for Fundamental Physics of the Universe, Via
Beirut 2, 34014, Trieste, Italy\\ $^{4}$Institute of Cosmology \& Gravitation,
University of Portsmouth, Burnaby Road, Portsmouth, UK \\ $^{5}$INFN --
Sezione di Trieste\\ $^{6}$Dipartimento di Fisica e Astronomia, Alma Mater
Studiorum Universit\`a di Bologna, via Piero Gobetti, 93/2, I-40129 Bologna,
Italy \\ $^{7}$INAF - Osservatorio di Astrofisica e Scienza dello Spazio, via
Piero Gobetti 93/3 1, I-40129 Bologna, Italy \\ $^{8}$INFN - Sezione di
Bologna, viale Berti Pichat 6/2, I-40127 Bologna, Italy }

\date{Accepted XXX. Received YYY; in original form ZZZ}

\pubyear{2019}

\begin{document}
\label{firstpage}
\pagerange{\pageref{firstpage}--\pageref{lastpage}}
\maketitle

\begin{abstract}
We present and test a new numerical method to determine the second--order
Lagrangian displacement field in the context of modified gravity theories. We
start from the extension of Lagrangian Perturbation Theory (LPT) to a class of
modifications of gravity, that can be described by a parametrized Poisson
equation with the introduction of a scale--dependent function. Then we exploit
fast Fourier transforms (FFTs) to compute the full source term of the
differential equation for the second--order Lagrangian displacement field. We
compare its mean to the source term computed for specific configurations for
which a $k$-dependent solution can be found numerically. We choose the
configuration that best matches the full source term, thus obtaining an
approximate factorization of the second--order displacement field as the space
term valid for standard gravity times a $k$-dependent, second--order growth
factor $D_2(k,t)$. This approximation is then used to compute second order
displacements for particles. The method is tested against N--body simulations
run with standard and $f(R)$ gravity: we rely on the results of a
friends-of-friends code run on the N--body snapshots to assign particles to
halos, then compute the halo power spectrum.  We find very consistent results
for the two gravity theories: second--order LPT (2LPT) allows to recover the
halo power spectrum of N--body simulations within $\sim 10\%$ precision to $k
\sim 0.2 - 0.4\ h\ {\rm Mpc}^{-1}$ (depending on the level of non-linearity),
as well as halo positions, with an error that is a fraction of the
inter--particle distance.  We show that, when considering the same level of
non--linearity in the density field, the performance of 2LPT with modified
gravity is the same (within $1\%$) as the one obtained for the standard
$\Lambda$CDM model with General Relativity.  When implemented in a computer
code, this formulation of 2LPT can quickly generate dark matter distributions
with $f(R)$ gravity, and can easily be extended to other modified gravity
theories, described in terms of a parametrized Poisson equation.
\end{abstract}


\section{Introduction}
Ever since the discovery of the accelerated expansion of the Universe
\citep{riess1998, perlmutter1999}, significant effort has been devoted to
trying to explain the mechanism behind it. Even though the standard
$\Lambda$CDM cosmological model successfully fits most observations on large
scales, the nature of Dark Energy is still one of the most challenging and
elusive open questions in cosmology and fundamental physics. Shedding light on
this topic is indeed a key target for future Large Scale Structure surveys,
such as Euclid \footnote{\url{https://www.euclid-ec.org/}}
\citep{laureijs2011}, DESI
\footnote{\url{https://www.desi.lbl.gov/}} \citep{levi2013}, LSST
\footnote{\url{https://www.lsst.org/}} \citep{abell2009} or
WFIRST \footnote{\url{https://wfirst.gsfc.nasa.gov/}}
\citep{spergel2013}. 

The $\Lambda$CDM model relies on the assumption that the growth of structures
in the Universe is driven by gravitational instability, described by
Einstein's General Relativity (hereafter GR).  Under this hypothesis, the
simplest interpretation for the gravitationally repulsive fluid responsible
for the cosmic accelerated expansion, and the only one that does not add new
degrees of freedom, is that of a cosmological constant $\Lambda$. Its natural
interpretation as the effect of vacuum energy poses strong theoretical
problems, such as fine tuning: the value of $\Lambda$ needed to explain the
recent accelerated expansion phase must be extremely small. This is in
contrast to the value predicted by quantum field theory, which is orders of
magnitude larger.  The cosmological constant problems have been extensively
discussed, see for example \citet{weinberg1989}, \citet{martin2012},
\citet{burgess2013}.

An alternative to the introduction of a cosmological constant to explain the
accelerated expansion is that General Relativity is not the correct theory for
gravity on cosmological scales.  Precision cosmology, that holds the promise
of providing accurate enough measurements to properly test different
scenarios, has prompted the development of a large number of modified gravity
models (hereafter MG, see for example \citet{joyce2015}, \citet{bull2016},
\citet{amendola2018}, \citet{ishak2019} for recent reviews on modified gravity
and cosmology).  Admittedly, General Relativity has succesfully passed all
tests up to now, from laboratory, to solar system, to the recent breakthroughs
provided by the observation of gravitational waves \citep{abbott2016} and the
imaging of the black hole in M87 \citep{eht2019}. As a consequence, any
alternative theory, in order to be viable, must satisfy very tight
constraints.  The proposed alternative models involve the introduction of an
additional fifth force which is opposed to gravity.  The behaviour of the
fifth force can be subdivided in three different regimes: on the largest
scales it must mimic $\Lambda$CDM, but with a large deviation from General
Relativity, in order to explain the accelerated expansion without the need of
a cosmological constant. On the smallest scales, the theory must reduce to GR:
to achieve this, a screening mechanism must be introduced. Finally, there
could still be deviations from GR on intermediate scales, where cosmological
observables carry specific signatures that can help disentangling between
different gravity theories.

Since possible signatures can be found in the mildly non--linear regime of
structure formation, it is of crucial importance that accurate theoretical
predictions are available, in order to compare to observations and place
constraints on different models.  The standard, and most reliable tools
employed to achieve this goal are N--body simulations.  However, full N--body
simulations are computationally expensive, and even more so if they are run
with modified gravity.  Their use becomes impractical, even in the standard GR
case, when addressing the computation of covariance matrices of observables
like the galaxy power spectrum or two--point correlation function; in this
case thousands of realizations are required to properly populate the matrices
and suppress the sampling noise.  For this reason, a variety of approximated
numerical methods have been developed, such as those implemented in {\sc
  pinocchio} (PINpointing Orbit Crossing Collapsed HIerarchical Objects,
\citealt{monaco2002-2, munari2017-2}), {\sc cola} \citep{tassev2013,
  izard2016, koda2016}, {\sc peak patch} \citep{bond1996, stein2019}, {\sc
  patchy} \citep{kitaura2014} and {\sc halogen} \citep{avila2015}. For a
recent review on approximated methods to generate halo catalogs, see
\citet{monaco2016}. These methods have been tested in the context of the
standard $\Lambda$CDM scenario (see \citealt{lippich2019}, \citealt{blot2019}
and \citealt{colavincenzo2019} for a comparison between different softwares).
Many of these methods are based on Lagrangian Perturbation Theory, so
extending them to MG theories requires to extend LPT first. This has been done
by several autors, like \citet{aviles2017}, and the extensions of the {\sc
  cola} approach to scalar-tensor modified gravity theories presented by
\citet{valogiannis2017} and \citet{winther2017}.  Recently, the {\sc
  pinocchio} code has been extended to massive neutrino cosmologies
\citep{rizzo2017}. That extension was based on the numerical result of
\cite{castorina2014} that the halo mass function in presence of massive
neutrinos can be obtained, with good accuracy, by using the the dark matter
(plus baryons) power spectrum, as if perturbations in the neutrino component
were always linear. The free streaming of massive neutrinos imprints a scale
depencence on the linear growth factor, $D_1 = D_1(k,t)$. This function was
obtained from the growth of the linear power spectrum as predicted by the {\sc
  camb} software \citep{lewis2002}, while the second-order growth rate was
obtained using the fit proposed by \cite{bouchet1995}, valid for GR in
$\Lambda$CDM model: $D_2 = -3/7\, D_1^2\Omega_m^{-1/143}$.  This approach was
adequate in the case of massive neutrinos, where the scale dependent growth is
due to the relativistic component but gravity is standard GR.  As we will show
later, in section~\ref{sec:test-lcdm-fit}, this simple technique does not give
accurate results when dealing with modifications of the gravity theory.

In this paper, we present and test a fast numerical method to compute 2LPT
displacements with a class of MG scalar--tensor theories, specializing it to
the case of $f(R)$ gravity. This is the first step toward a full extension of
the {\sc pinocchio} code.  The main problem to face is the fact that, unlike
in the case of standard GR, the LPT displacement terms cannot be factorized
into space-- and time--dependent functions.  At second-order this leads to a
very complicated integro--differential equation, whose numerical solution is
very hard to obtain. \citealt{winther2017} already proposed an approximate way
to achieve a factorization into a space--dependent part and a mildly
scale--dependent growth factor $D_2(k, t)$. With respect to that work, we
quantify the error made by approximating the full source term of the equation
of the 2LPT displacement potential, and investigate the effect of this error
by predicting the non--linear power spectrum of dark matter halos and
comparing to the one measured from the output of an N--body simulation run
with {\sc mg--gadget} \citep{puchwein2013} with $f(R)$ gravity.

The paper is structured as follows: in section \ref{sec:lpt} we give an
overview of LPT for the standard $\Lambda$CDM model. In section \ref{sec:mg}
we summarize the equations used to extend LPT to scalar--tensor theories, and
introduce the $f(R)$ modified gravity model we are considering.  In section
\ref{sec:method} we describe a new numerical method that allows to compute the
full source term of the second order differential equation for the
displacement field. This allows to test different configurations in order to
find the one that best matches the full solution, as well as to quantify the
error introduced by approximating the second order growth factor. We perform a
specific test by comparing to the outputs of a full N--body simulation,
presented in section \ref{sec:test}, to validate our method.  In section
\ref{sec:test-lcdm-fit} we also test the approximation proposed by
\citealt{bouchet1995} to compute the second--order growth factor from
$D_1(k,a)$, showing that this approach is not suitable in the case of modified
gravity.  In section \ref{sec:conclusion} we draw our conclusions and discuss
future works.
%
%
\section{Theoretical Framework}
\label{sec:theory}
\subsection{Lagrangian Perturbation Theory in $\Lambda$CDM}
\label{sec:lpt}
Lagrangian Perturbation Theory, pioneered by Zel'dovich
(\citet{zeldovich1970}, see \citet{bouchet1996} for a review), has proven a
very powerful tool and is indeed the foundation on which many approximated
methods rely.  It is based on a Lagrangian description of the dynamics of
cosmic fluids, following particles' trajectories instead of studying the
evolution of the density and velocity fields in a fixed frame as in Eulerian
perturbation theory. It can be seen as a coordinate change, with the main
quantity being the displacement field $\vec{\Psi}$ which maps the initial
position $\vec{q}$ of a fluid element to the final, Eulerian position
$\vec{x}$ through
\begin{equation}
\label{eq:position}
\vec{x}(\vec{q},a) = \vec{q} + \vec{\Psi}(\vec{q},a) \; , 
\end{equation}
where $a$ is the scale factor. As long as the displacement is small, it can be
expanded in a perturbation series; moreover, as long as $\vec{\Psi}$ is curl
free (since it is second--order), it can be written as the gradient of a
scalar potential $\phi$:
\begin{equation}
\label{eq:psigradphi}
\vec{\Psi}(\vec{q},a) = \nabla_{\vec{q}} \; \phi(\vec{q},a) \; ,
\end{equation}
with $\nabla_{\vec{q}} = \partial / \partial \vec{q}$ being the gradient in
Lagrangian coordinates.  The equation of motion for the particle trajectory
can be written as
\begin{equation}
\label{eq:motion}
a^2 H^2(a) \left[ \frac{\deriv^2}{\deriv a^2} + \left( \frac{3}{a} +
  \frac{H'(a)}{H(a)} \frac{\deriv}{\deriv a} \right) \right] \vec{x} = a^2 H^2
\hat{T} \vec{x} = - \nabla_{x} \Phi_N \; ,
\end{equation}
where the $'$ denotes derivation with respect to the scale factor, $H(a)$ is
the Hubble parameter, $\Phi_N$ is the gravitational potential and we defined
the $\hat{T}$ operator as the quantity between square brackets in
eq. \ref{eq:motion}. Note that here $\nabla_{\vec{x}}=\partial / \partial
\vec{x}$ is the gradient in Eulerian coordinates. By imposing matter
conservation, it is possible to write the relation between the displacement
field and the overdensity $\delta$:
\begin{equation}
\label{eq:deltajacobian}
\delta(\vec{x},a) = \frac{1- J(\vec{q},a)}{J(\vec{q},a)} \; ,
\end{equation}
where $J(\vec{q},a)$ is the determinant of the Jacobian of the transformation:
\begin{equation}
J_{ij} = \frac{\partial x^i}{\partial q^j} = \delta_{ij} + \frac{\partial
  \Psi^i}{\partial q^j} \; .
\end{equation}
By taking the divergence of eq. \ref{eq:motion} together with the Poisson
equation and eq. \ref{eq:deltajacobian} we can write the evolution equations
for the first and second order Lagrangian potentials:
\begin{equation}
\begin{aligned}
&a^2 H^2 \left( \hat{T} - 4 \pi G \bar{\rho} \right) \phi_{,ii}\first = 0 \; ,
  \\ &a^2 H^2 \left( \hat{T} - 4 \pi G \bar{\rho} \right) \phi_{,ii} \second =
  - 4 \pi G \bar{\rho} \left[ \frac{1}{2} \left( \phi_{,ii} \phi_{jj} -
  \phi_{,ij} \phi_{,ji} \right) \right] \; .
\end{aligned}
\end{equation}
Here $,i$ denotes the derivative with respect to $q_i$, and we adopt the
standard notation of summation over repeated indices. Since the operator
acting on $\phi \first$ and $\phi \second$ is only a function of time, the
time evolution can be factored out and the potentials can be written as the
(time--dependent) growth factors times the initial potentials:
\begin{equation}
\begin{aligned}
&\vec{\phi}\first(\vec{q},a)  = D_1(a) \vec{\phi}\first(\vec{q},a_{in}) \; ,\\
&\vec{\phi}\second(\vec{q},a) = D_2(a) \vec{\phi}\second(\vec{q},a_{in}) \; .
\end{aligned}
\end{equation}
Given an initial displacement field, the computation of potentials and
displacements for any time is thus straightforward, once the equation for the
first and second order growth factors are solved:
\begin{equation}
\begin{aligned}
&a^2 H^2 \left( \hat{T} - 4 \pi G \bar{\rho} \right) D_1(a) = 0 \; ,\\
&a^2 H^2 \left( \hat{T} - 4 \pi G \bar{\rho} \right) D_2(a) = - 4 \pi G
\bar{\rho} D_1^2(a) \; ,
\end{aligned}
\end{equation}
The initial, first order potential is directly linked to the density field
through eq. \ref{eq:deltajacobian}:
\begin{equation}
\phi \first_{,ii}(\vec{q}, a_{in}) = - \delta \first(\vec{q}, a_{in}) \; ,
\end{equation}
while the second order can be written as
\begin{equation}
\label{eq:second-ini}
\phi \second_{,ii} (\vec{q}, a_{in}) = \frac{1}{2} \left[ \phi \first_{,ii} \phi
  \first_{,jj} - \phi \first_{,ij} \phi \first_{,ji} \right] (\vec{q}, a_{in}) \; ,
\end{equation} 
and can be easily and readily computed with Fast Fourier Transorms (FFTs) from
the initial first order Lagrangian potential $\phi \first(\vec{q},a_{in})$.

The possibility to factor out the time evolution to compute displacements in
the particles' positions is the key feature that makes this approach ideal to
be implemented in fast, approximated methods that simulate the formation of
the Large Scale Structure of the Universe. However, as anticipated above and
described in detail in section \ref{sec:lpt-mg}, one effect of modified
gravity is that the growth factors become scale dependent. As a consequence,
separating out the time evolution to compute displacements at any given time
is not possible anymore, and both the theoretical and computational treatment
of LPT with modified gravity become more involved.
\subsection{Modified Gravity}
\label{sec:mg}
In this work we focus on MG models that mimic $\Lambda$CDM on large scales,
while on intermediate scales they include a fifth force that is due to a new
scalar degree of freedom.  Moreover, they need a screening mechanism to comply
with tight Solar System constraints at small scales.  The fifth force caused
by the gravity modification introduces mode coupling even at the linear level;
additionally, in order to properly describe non--linear scales, the
Klein--Gordon equation for the scalar field must be solved.  Following the
approach of \citet{koyama2009}, the modified Poisson equation and the
Klein--Gordon equation can be written as:
\begin{equation}
\label{eq:modifiedpoisson}
\frac{1}{a^2} \nabla^2 \Phi = 4 \pi G \bar{\rho} \delta - \frac{1}{2a^2}
\nabla^2 \varphi \; ,
\end{equation}
\begin{equation}
\label{eq:kg}
(3 + 2 \omega_{BD}) \frac{1}{a^2} \nabla^2 \varphi = -8 \pi G \bar{\rho}
\delta + \mathrm{NL} \; ,
\end{equation}
where $\Phi$ is the gravitational potential, $\bar{\rho}$ is the background
matter density, $\varphi$ is the scalar field that encodes the modification of
gravity, $\omega_{BD}$ is the Brans--Dicke parameter, and {\it NL} are
possible non--linearities that might arise in the Lagrangian. Going to Fourier
space, eq. \ref{eq:kg} can be written as:
\begin{equation}
\label{eq:kgfourier}
(3 + \omega_{BD}) \frac{k^2}{a^2} \varphi_k = 8 \pi G \bar{\rho} \delta_k -
\mathcal{I}(\varphi_k) \; .
\end{equation}
The term $\mathcal{I}(\varphi_k)$ is the scalar field self--interaction, that
is related to the screening mechanism responsible of recovering GR on small
scales. It can be expanded as $\mathcal{I}(\varphi_k) = M_1(k,a) \varphi_k +
\delta \mathcal{I}(\varphi_k)$, with
\begin{equation}
\label{eq:nlselfint}
\begin{aligned}
\delta \mathcal{I} (\varphi_k) = \frac{1}{2} \int \frac{\deriv^3 k_1 \deriv^3
  k_2}{(2 \pi)^3} &\delta_D(\vec{k} - \vec{k}_{12}) M_2(\vec{k}_1, \vec{k}_2,a) \\
&\times \varphi(\vec{k}_1,a) \varphi(\vec{k}_2,a) + \mathcal{O}(\varphi_k^3) \; ,
\end{aligned}
\end{equation}
where the $M_i$ functions are in general scale and time dependent and their
functional form depends on the particular model considered. In the following
section we will focus on scalar--tensor theories of gravity, targeting in
particular the $f(R)$ family of gravity models (see \citet{defelice2010} for a
review). Our method however is general, and can be applied to other
scalar--tensor theories, provided that the MG potential can be split in a
background value plus perturbations, and the perturbations can be Taylor
expanded (see eq. \ref{eq:mg-pot-exp} below).
\subsection{f(R) gravity}
\label{sec:fr}
In $f(R)$ models the Einstein--Hilbert Lagrangian density is modified to include
a function of the Ricci scalar $R$:
\begin{equation}
\label{eq:fr-lag}
\mathcal{L}_R = \sqrt{-g} \left(R + f(R) \right) \; .
\end{equation}
This possible extension to General Relativity has been widely developed, both
in terms of theoretical predictions and possible observational signatures. The
functional form of $f(R)$ is bounded by the requirement of satisfying Solar
System constraints and reproducing the $\Lambda$CDM expansion history; several
functional forms meet these requirements.  The one we are considering in this
paper is that described in \citet{hu2007}.  While constraints on model
parameters are getting tighter and tighter, particular effort has recently
been put into investigating them in light of the degeneracy with the mass of
neutrinos (see e.g. \citet{baldi2014}, \citet{hu2015}, \citet{giocoli2018},
\citet{wright2019}).

By varying the action constructed with the modified Lagrangian of
eq. \ref{eq:fr-lag} with respect to the metric, and then taking the trace of
the resulting field equations, one obtains:
\begin{equation}
3 \mathlarger{\mathlarger{\square}} f_R = R ( 1 - f_R) + 2f - 8 \pi G \rho \; ,
\end{equation}
where $f_R = \deriv f(R) / \deriv R$. Equivalently, one can split $f_R$ and
$R$ in background quantities plus perturbations $\delta f_R$ and $\delta
R$. In the quasi--static approximation one can write:
\begin{equation}
\frac{3}{a^2} \nabla^2 \delta f_R = - 8 \pi G \bar{\rho} \delta + \delta R \; ,
\end{equation}
which is nothing but the Klein--Gordon equation for a scalar field with
potential $\delta R$ and Brans--Dicke parameter $\omega_{BD} = 0$. The
potential can be expanded as
\begin{equation}
\label{eq:mg-pot-exp}
\delta R = \sum_k \frac{1}{k!} M_k (\delta f_R)^k \qquad , \qquad M_k =
\left. \frac{\deriv^k R(f_R)}{\deriv f_R^k} \right\rvert_{f_R = \bar{f}_R}
\end{equation}
For $f(R)$ gravity the coefficients $M_k$ only depend on time; this is an
important feature to the approach we propose in this work (described in
section \ref{sec:method}). In the following treatment we will consider
Hu--Sawicki $f(R)$, for which we have:
\begin{equation}
\label{eq:husawicki}
f(R) = - \beta^2 \frac{c_1 (R/\beta^2)^n}{c_2 (R/\beta^2)^n + 1} \; , 
\end{equation}
where $\beta^2$ is the mass scale, defined as $\beta^2 = H_0^2 \Omega_{m,0}$,
and $c_1$,$c_2$ and $n$ are free parameters of the model. The model is
consistent with a $\Lambda$CDM background expansion if one chooses $c_1/c_2 =
6 \Omega_{\Lambda} / \Omega_{m,0}$, thus leaving only two free parameters that
can be recast in terms of the value of $f_R$ today, $f_{R0}$, and $n$. By
fixing $n=1$, the $M_k$ coefficients can be written as:
\begin{equation}
\begin{aligned}
\label{eq:mihusawicki}
&M_1(a) = \frac{3}{2} \frac{H_0^2}{\lvert f_{R0} \rvert} \frac{(\Omega_{m,0}
  a^{-3} + 4 \Omega_{\Lambda})^3}{(\Omega_{m,0} + 4
  \Omega_{\Lambda})^2} \; ,\\
&M_2(a) = \frac{9}{4} \frac{H_0^2}{\lvert f_{R0}
  \rvert^2} \frac{(\Omega_{m,0} a^{-3} + 4 \Omega_{\Lambda})^5}{(\Omega_{m,0}
  + 4 \Omega_{\Lambda})^4} \; .
\end{aligned}
\end{equation}
\subsection{LPT with modified gravity}
\label{sec:lpt-mg}
A proper formulation of LPT in the framework of scalar--tensor modified
gravity theories has been proposed only recently (see \citealt{aviles2017},
\citealt{valogiannis2017}, \citealt{winther2017}).  For a full theoretical
description we refer to \citet{aviles2017}, where a general formalism to
compute Lagrangian displacement fields with MG up to third order was
presented; here we report just the basic equations necessary to describe our
method.

\begin{figure}
  \centering
  \includegraphics[width=\columnwidth]{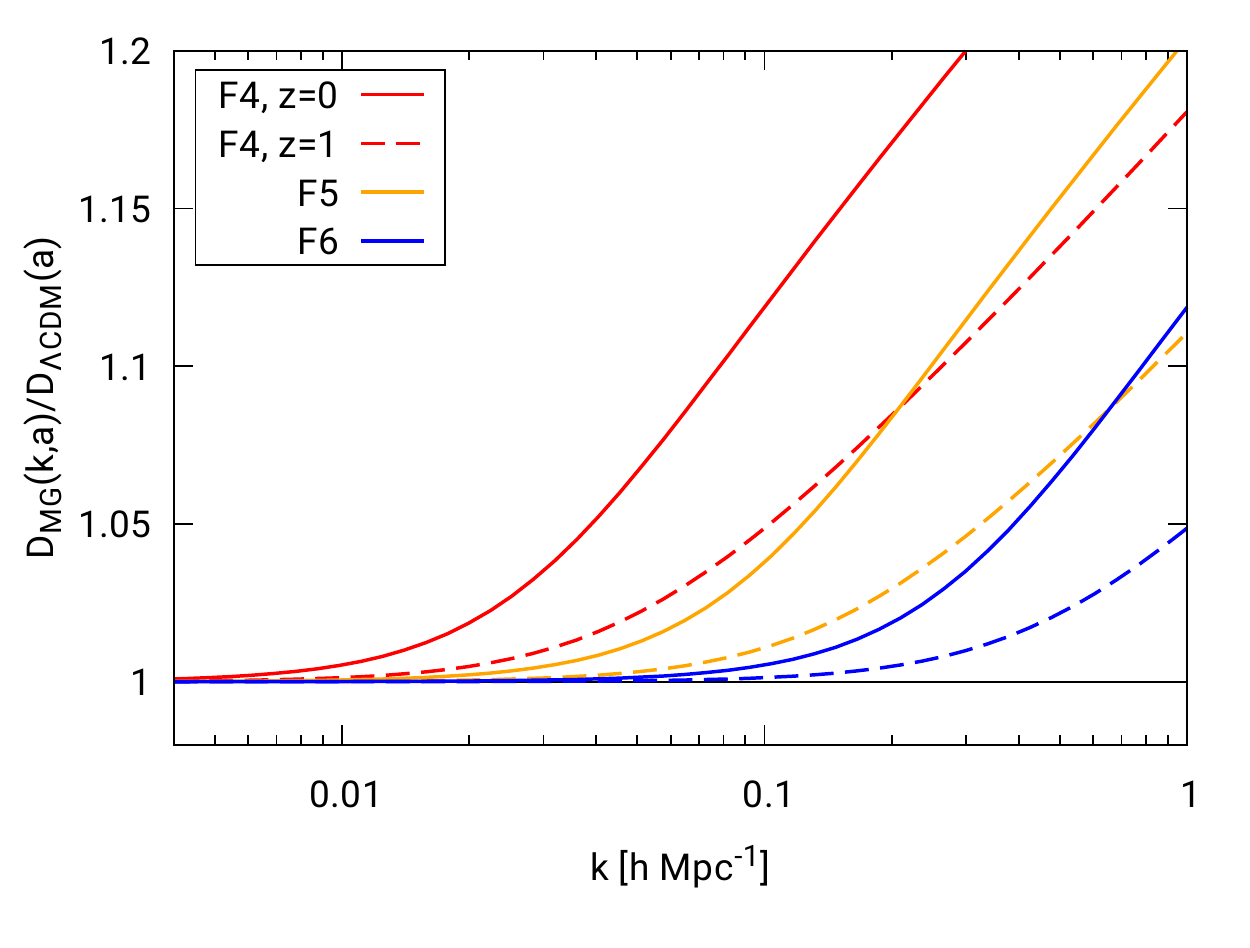}
  \caption{Solution to eq. \ref{eq:d1} for Hu--Sawicki $f(R)$ with n=1
    normalized to the $\Lambda$CDM linear growing mode for three different
    values of the $f_{R0}$ parameter ($f_{R0}=-10^{-4}$ in red, labelled as
    F4; $f_{R0}=-10^{-5}$ in orange, labelled as F5; $f_{R0}=-10^{-6}$ in
    blue, labelled as F6), shown for $z=0$ (solid line) and $z=1$ (dashed
    line).}
  \label{fig:d1}
\end{figure}

By substituting eq. \ref{eq:kgfourier} in the Fourier space version of the
modified Poisson equation \ref{eq:modifiedpoisson}, and then combining with
the equation of motion \ref{eq:motion}, we can write the evolution equation
for the first order displacement field in Fourier space as:
\begin{equation}
\label{eq:first-displ-mg}
a^2 H^2 (\hat{T} - 4 \pi G \bar{\rho} \mu(k,a)) \; \fourier [ \phi
  \first_{,ii}](\vec{k},a) = 0 \; ,
\end{equation}
where $\fourier$ is the Fourier transform operator, and
\begin{equation}
\mu(k,a) = 1 + \frac{1}{3} \frac{k^2/a^2}{k^2/a^2 + m^2(a)} \; .
\end{equation}
The $m^2(a)$ function represents the mass of the scalar field, and is related
to $M_1(a)$ by $M_1(a) = 3 m^2(a)$.  It is no longer possible to separate time
and space, since the operator acting on the first order displacement potential
is no longer time--dependent only, due to the presence of $\mu(k,a)$ in
eq. \ref{eq:first-displ-mg}. Nonetheless, we can separate time for each
Fourier mode, so that:
\begin{equation}
\fourier [\phi \first_{,ii}](\vec{k},a) = D_1(k,a) \; \fourier [\phi
  \first_{,ii}] (\vec{k}, a_{in}) \; ,
\end{equation}
where $D_1(k,a)$ is the solution of:
\begin{equation}
\label{eq:d1}
a^2 H^2 (\hat{T} - 4 \pi G \bar{\rho} \mu(k,a)) D_1(k,a) = 0 \; .
\end{equation}
We note that the first order growth factor is now scale dependent, due to the
presence of the $\mu(k,a)$ function in the differential equation.  However,
the scale dependence is fully enclosed in $\mu$, and is only related to the
modulus of $k$. The linear growth factor can then be computed by fixing a
value for $k$ and solving the differential equation, then repeating for a set
of $k$--values and finally interpolating to obtain the function at any $k$. We
numerically solve eq. \ref{eq:d1} with a standard Runge--Kutta algorithm, with
initial conditions for $D_1(k,a)$ set to the growing mode for a matter
dominated (Einstein--de Sitter) Universe, namely $D_1(a_{in}) = a_{in}$ and
$D_1'(a) \lvert_{a=a_{in}} = 1$.  The resulting linear growth factor is then
normalized so that $D_1(k=0, a=1) = 1$. The result is shown in
fig. \ref{fig:d1}, where we plot the ratio between the MG linear growth factor
$D_{MG}$ and the $\Lambda$CDM one in the case of $n=1$ Hu--Sawicki $f(R)$, for
three different values of the $f_{R0}$ parameter and two different redshifts.
Once again, the initial first--order displacement field can be determined from
the initial density field, and its evolution computed my multiplying it by
$D_1(k,a)$. However, when going to second order this kind of separation cannot
be done; the second order growth factor now depends on three wavenumbers $k$,
$k_1$ and $k_2$ and on the dot product $\vec{k}_1 \cdot \vec{k}_2$.  The second
order displacement field can be written (in Fourier space) as an integral over
$k_1$ and $k_2$:
\begin{equation}
\label{eq:displ2}
\fourier [\phi \second_{,ii}](\vec{k},a) = \int \frac{\deriv^3 k_1 \deriv^3
  k_2}{(2 \pi)^3} \delta_D(\vec{k} - \vec{k}_{12})
D_2(k,k_1,k_2,a) \delta_1 \delta_2 \; ,
\end{equation}
where $\delta_D$ is the Dirac's delta, $\vec{k}_{12} = \vec{k}_1 + \vec{k}_2$,
$\delta_i = \delta(\vec{k}_i)$ is the linear density contrast evaluated at
present time and $D_2(k, k_1, k_2, a)$ is the
scale--dependent second order growth rate obtained by solving (see
\citealt{aviles2017}, where a full derivation of the following equation can be
found):
\begin{equation}
\label{eq:full-d2}
\begin{aligned}
a^2 H^2(a) &\left[\hat{T} - 4 \pi G \bar{\rho} \mu(k) \right]
D_2(k,k_1,k_2,a)\\ &= 4 \pi G \bar{\rho} D_1(k_1,a) D_1(k_2,a) \Bigg\{ \mu(k)
+ \\ & - \frac{( \vec{k_1} \cdot \vec{k_2})^2}{k_1^2 k_2^2} \left[ \mu(k_1) +
  \mu(k_2) - \mu(k) \right] + \\ & + \frac{m^2(a)}{\Pi(k)} \left[2
  \frac{(\vec{k_1} \cdot \vec{k_2})^2}{k_1^2 k_2^2} \left( \mu(k_1) + \mu(k_2)
  - 2 \right) + \right.\\ &\left. \frac{\vec{k_1} \cdot \vec{k_2}}{k_1^2}
  (\mu(k_1) - 1) + \frac{\vec{k_1} \cdot \vec{k_2}}{k_2^2} (\mu(k_2) - 1)
  \right] + \\ & - \frac{2}{27} \, 4 \pi G \bar{\rho} \, \frac{k^2}{a^2} \,
\frac{M_2(a)}{\Pi(k) \Pi(k_1) \Pi(k_2)} \Bigg\}\; .
\end{aligned}
\end{equation}
Here we did not explicity write the time dependence of $\mu(k,a)$ and
$\Pi(k,a)$ in the previous equation to simplify it.  

The presence of the Dirac's delta in eq. \ref{eq:displ2} requires that
$\vec{k} = \vec{k}_1 + \vec{k}_2$, so that the integral runs over all possible
triangle configurations formed by $\vec{k}_1$, $\vec{k}_2$ and $\vec{k}$ in
Fourier space.  Because of this, implementing the full solution for the second
order displacements would require to solve a different equation for each
wavenumber $\vec{k}$, whose source term includes a 9--dimensional
integral. While not unfeasible in principle, this computation would be very
time consuming, making 2LPT a poor alternative to full N--body simulations.

One possible alternative, already explored by \citealt{winther2017}, is to
find an approximation for $D_2(k,a)$, in order to achieve an effective
factorization of the second order--potential into the same space part as in GR
(to be computed with Fast Fourier Transforms) and an effective $k$--dependent
growth rate:
\begin{equation}
\phi \second (\vec{k},a) = D_2(k,a) \phi \second(\vec{k},a_{in})
\end{equation}
In particular, one can choose a triangle configuration for $\vec{k}$,
$\vec{k}_1$ and $\vec{k}_2$, solve eq. \ref{eq:full-d2} to find
$D_2(k,k_1,k_2,a)$ and then compute the displacement field in the standard
way, with $\phi \second(\vec{k},a_{in})$ being the Fourier--space version of
the initial second order displacement field of eq. \ref{eq:second-ini}.
%
%
\section{Method: approximating the 2LPT displacement field}
\label{sec:method}
As discussed in the previous section, our goal is to find an approximation for
the second--order growth rate which allows to readily compute the second order
displacement field.  Moreover, we want to quantify the deviation of the
approximation from the full solution.  Our approach is to compute the full
source term of the differential equation for the 2LPT displacement field by
taking advantage of FFTs, and then compare it to analytical expressions for
different triangle configurations, in order to find the one that best matches
the full source term.  Next, we numerically solve the differential equation for
$D_2$ for the chosen triangle configuration, and use it to approximate the
evolution of the displacement field.

The second order displacement field in general, scalar--tensor theories of
gravity (where the scalar field potential can be expanded as in
eq. \ref{eq:mg-pot-exp}) is the solution of eq. \ref{eq:displ2}.  The growth
factor can be computed by solving eq. \ref{eq:full-d2}.  This equation reduces
to the standard, $\Lambda$CDM one for $\mu(k,a)=1$.  The dependence on closed
triacles in Fourier space is related to the presence of derivatives of the
first--order displacement field as well as the $M_k$ functions, which can in
principle bear a scale dependence.  In the special case of $f(R)$ gravity
theories, the $M_k$ functions only depend on time, so they can be taken out of
the integral we need to solve to compute $\phi \second (\vec{k},
a)$. Eq. \ref{eq:displ2} can then be written by expressing the Fourier--space
integrals as Fourier transforms of local, non--linear functions in
real--space.  It is then possible to take advantage of FFTs to compute the
full source term of the differential equation. The validity of this approach
is not limited to $f(R)$ models but extends to all theories where the MG
scalar potential can be expanded into scale independent coefficients.  The
full equation for 2LPT displacements can be written as:
\begin{equation}
\label{eq:full}
a^2 H^2 (\hat{T} - 4 \pi G \bar{\rho} \mu(k,a)) \; \fourier [ \phi
  \second_{,ii}](\vec{k},a) = S_1 + S_2 + S_3 + S_4 \; ,
\end{equation}
where
\begin{equation}
\label{eq:s1}
S_1 = 4 \pi G \bar{\rho} \; \fourier \left[ \phi \first_{,ij} \; \fourier^{-1}
  \left[\mu(k,a) \; \fourier[\phi \first_{,ji}] \right] \right] \; ,
\end{equation}
\begin{equation}
\label{eq:s2}
S_2 = -2 \pi G \bar{\rho} \mu(k,a) \; \fourier \left[ \phi \first_{,ii} \phi
  \first_{,jj} - \phi \first_{,ij} \phi \first_{,ji} \right] \; ,
\end{equation}
\begin{equation}
\label{eq:s3}
S_3 = \left( \frac{8 \pi G \bar{\rho}}{3} \right)^2 \frac{M_2(a)}{12}
\frac{k^2/a^2}{\Pi(k,a)} \; \fourier \left[ \left(\fourier^{-1} \left[ \frac{\delta
      \first_k}{\Pi(k,a)} \right] \right)^2\right] \; ,
\end{equation}
\begin{equation}
\label{eq:s4}
\begin{aligned}
S_4 = - \frac{8 \pi G \bar{\rho}}{3} \frac{m^2(a)}{2 a^2} &\frac{1}{\Pi(k,a)}
\; \fourier \left[ 2 \phi \first_{,ij} \left( \fourier^{-1} \left[
    \frac{\delta \first _k}{\Pi(k,a)} \right] \right)_{,ij}
  \right. \\ &\left. + \phi \first_{,iij} \left( \fourier^{-1} \left[
    \frac{\delta \first _k}{\Pi(k,a)} \right] \right)_{,j}\right] \; .
\end{aligned}
\end{equation}
Here $\Pi(k,a) = k^2 / a^2 + m^2(a)$ and the $\phi \first$, $\delta \first$
fields are evolved with the linear scale--dependent growth factor $D_1(k,a)$.
The $S_1$ and $S_2$ terms come from keeping second order terms in the Poisson
equation and the equation of motion.  The $S_3$ term is related to the
second--order scalar field self--interaction (NL in eq. \ref{eq:kg}).
 Finally, the $S_4$ term (first introduced by \citealt{aviles2017}), is a
geometric term, due to the fact that we are performing Fourier transforms in
Lagrangian Fourier space, not Eulerian.

\begin{figure}
  \centering
  \includegraphics[width=\columnwidth]{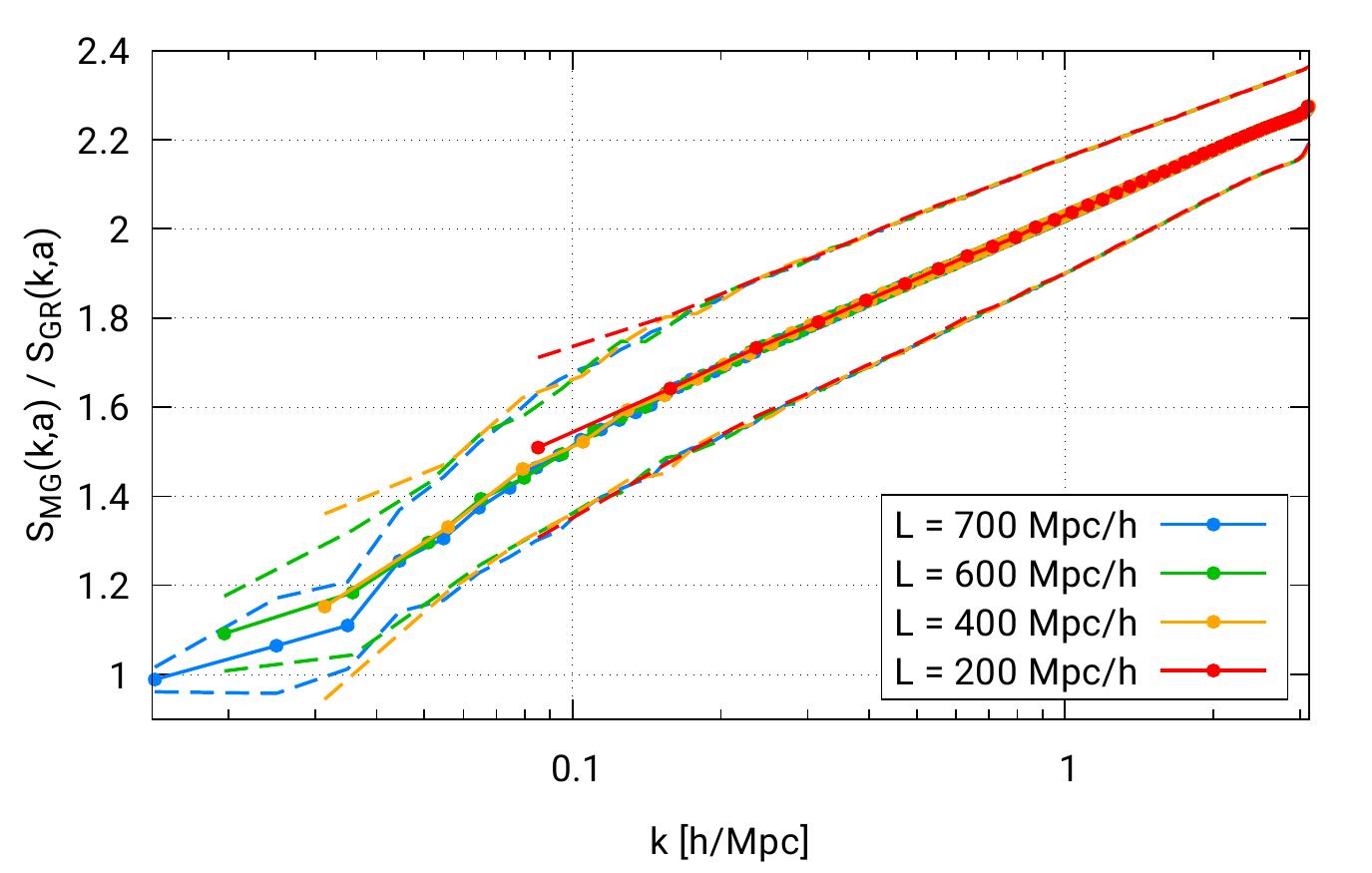}
  \caption{Source term of the second--order displacement field differential
    equation with $f(R)$ modified gravity, normalized to the one in GR at
    $z=0$. The modified gravity model is $n=1$ Hu--Sawicki with
    $f_{R0}=-10^{-4}$. Different colors correspond to different box sizes, all
    with the same resolution of $1$ particle per $\mathrm{Mpc} \;
    h^{-1}$. Each solid line is the result binned in $k$; dashed lines
    represent $1\sigma$ deviation from the mean value within each $k$--bin.}
  \label{fig:source-term}
\end{figure}

\begin{figure}
  \centering
  \includegraphics[width=\columnwidth]{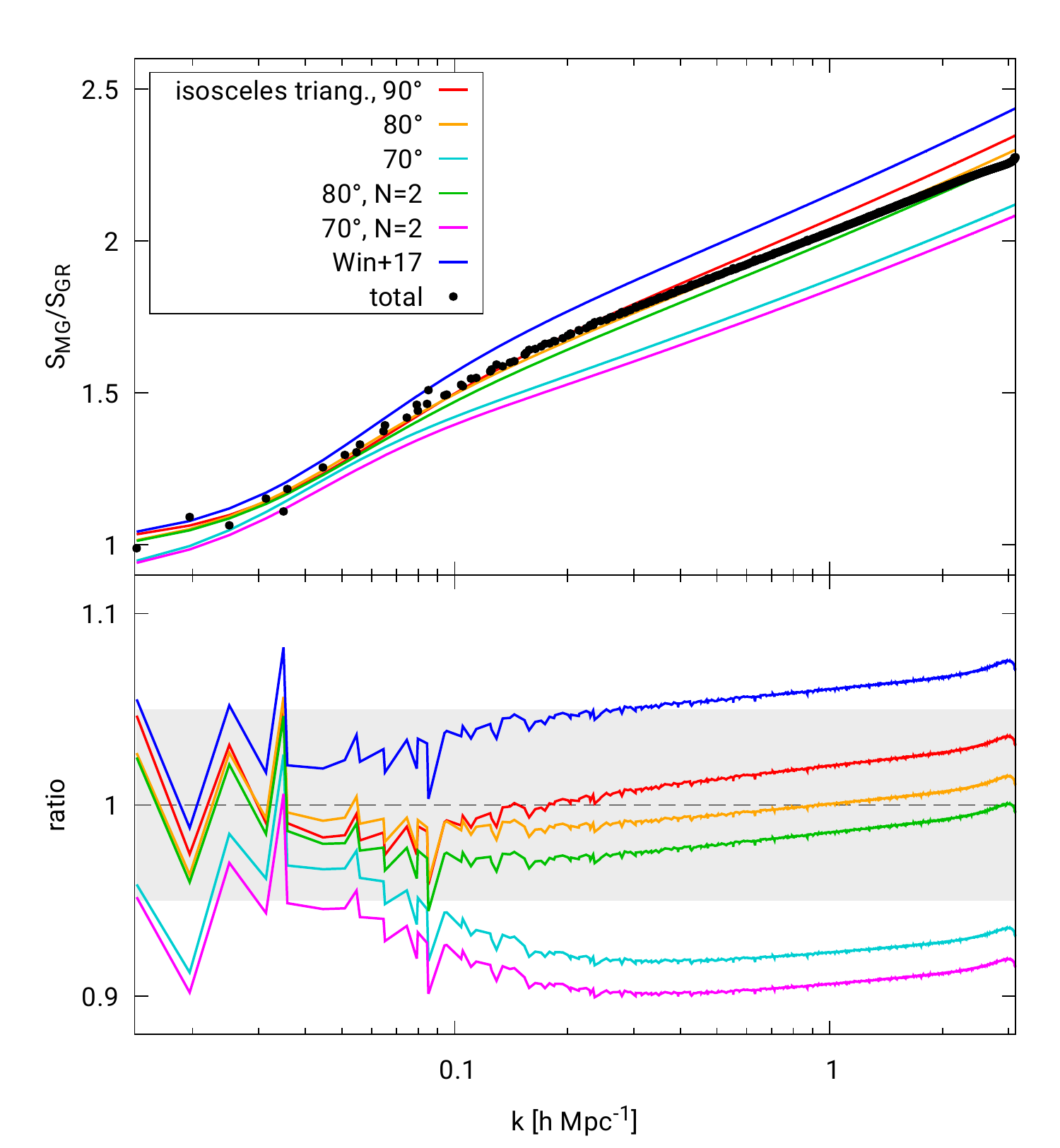}
  \caption{Top panel: comparison of the full source term (black dots) to
    different triangle configurations at redshift $z=0$. The red, orange and
    cyan lines represent isosceles triangles respectively with angle
    $90^{\circ}$ (orthogonal configuration), $80^{\circ}$ and $70^{\circ}$
    between $k_1$ and $k_2$. The green and magenta lines represent triangles
    with $k_1=2k_2$ and angle $80^{\circ}$ and $70^{\circ}$ respectively
    between $k_1$ and $k_2$. In blue is shown the approximation adopted in
    \citet{winther2017}. Bottom panel: ratio of the full solution to different
    configurations. The grey shaded area represents a $5\%$ deviation from the
    full source term.}
  \label{fig:triang-ratio}
\end{figure}

The method we adopt is the following: we generate a linear density field on a
regular grid, we compute the first order growth factor $D_1(k,a)$ by
numerically solving eq. \ref{eq:d1}, then use it to evolve the field.  Next we
compute the $S_i$ terms of eq. \ref{eq:full}, going back and forth from
Fourier space to configuration space to solve the integrals. We divide the
source term by the equivalent quantity evaluated for $\Lambda$CDM. The result
is a quantity that depends on $\vec{k}$, which we bin in a grid of
$k$--values, computing its average and scatter within each bin. Then we
compare this average with the analytical expressions obtained using various
triangle configurations in Fourier space.  The result is shown in
fig. \ref{fig:source-term}, where we show the computation of the full source
term of the differential equation divided by its equivalent evaluated for a
$\Lambda$CDM cosmology, at z=0. The solid lines represent the source term for
boxes with different sizes ($200 \; \mathrm{Mpc} \; h^{-1}$, $400 \;
\mathrm{Mpc} \; h^{-1}$, $600 \; \mathrm{Mpc} \; h^{-1}$, $700 \; \mathrm{Mpc}
\; h^{-1}$) with a fixed resolution of 1 particle / $\mathrm{Mpc} \; h^{-1}$.
For each box we produce two realizations, one with modified gravity and one
with standard general relativity, both with the same initial conditions in
order to have the same modes and sample variance.  We then compute the ratio
of the two and compute average and standard deviation in bins of $k$. Dashed
lines show the obtained $1\sigma$ standard deviation of the distribution of
the points in each bin: this represents the scatter, due to the fact that the
source term depends on the vector $\vec{k}$.  This scatter provides a measure
of how accurate a factorization in terms of a mildly $k$--dependent growth
rate $D_2(k,t)$ is: even though the source term is not completely separable,
the standard deviation is always below $\sim 0.2$, and goes to zero at large
scales, as expected.  Moreover, the average varies smoothly with $k$, and the
standard deviation of the mean within each bin is not large, $\sigma /
\sqrt{N} \sim 10^{-6}$ (with N the number of wavemodes in each bin).  We can
conclude that the average is measured with a good precision, and can be used
to the purpose of finding an approximation to $D_2$.
We then compare the average ratio of source term to the same quantity,
obtained analytically by adopting different triangle configurations: the
result is shown in fig. \ref{fig:triang-ratio}. The top panel shows the full
source term (divided by the GR one) of fig. \ref{fig:source-term} with black
dots, and different triangle configurations (solid lines), while in the bottom
panel we show the percent difference between the full source term and
different triangle configurations.  First we compare to orthogonal ($k_1=k_2$,
$\theta = 90^{\circ}$), equilateral ($k_1=k_2$, $\theta=60^{\circ}$) and
squeezed ($k_1 \simeq 0$, $k_2=k$) configurations.  We find the solution to be
very close to the orthogonal configuration, and above the equilateral
one. These are both isosceles triangles with $k_1=k_2$ and angle between
$\vec{k}_1$ and $\vec{k}_2$ respectively $\theta=90^{\circ}$ and
$\theta=60^{\circ}$.  We therefore focus on isosceles triangles, keeping
$k_1=k_2$ and varying the angle. We find the best configuration to be the
orthogonal one (red line in fig. \ref{fig:triang-ratio}, hereafter T1) and the
one with $\theta=80^{\circ}$ (orange line in fig. \ref{fig:triang-ratio},
hereafter T2).  We find that both T1 and T2 give results that are well within
$1 \%$ with respect to the full source term, in particular for the mildly
intermediate scales we are interested in describing with 2LPT.  We also
compare the source term to triangle configurations with different ratio
$k_1/k_2$ and fixed angle $80^{\circ}$, finding that increasing the ratio
$k_1/k_2$ gives a worse match to the source term (green and magenta lines of
fig.\ref{fig:triang-ratio}).  The approximation proposed by
\citealt{winther2017}, is shown in blue in fig. \ref{fig:triang-ratio}, and
corresponds to fixing $k_1=k_2$, $\theta=90^{\circ}$ in eq. \ref{eq:full-d2},
but the first order growth rates in that equation are computed as $D_1(k)$
instead of $D_1(k_1)$, $D_1(k_2)$.  This choice gives a slight overestimation
of the source term, but the deviation is still within $5\%$ up to $k \sim 0.2
h \, \mathrm{Mpc}^{-1}$.

\begin{figure*}
\centering
  \includegraphics[width=2\columnwidth]{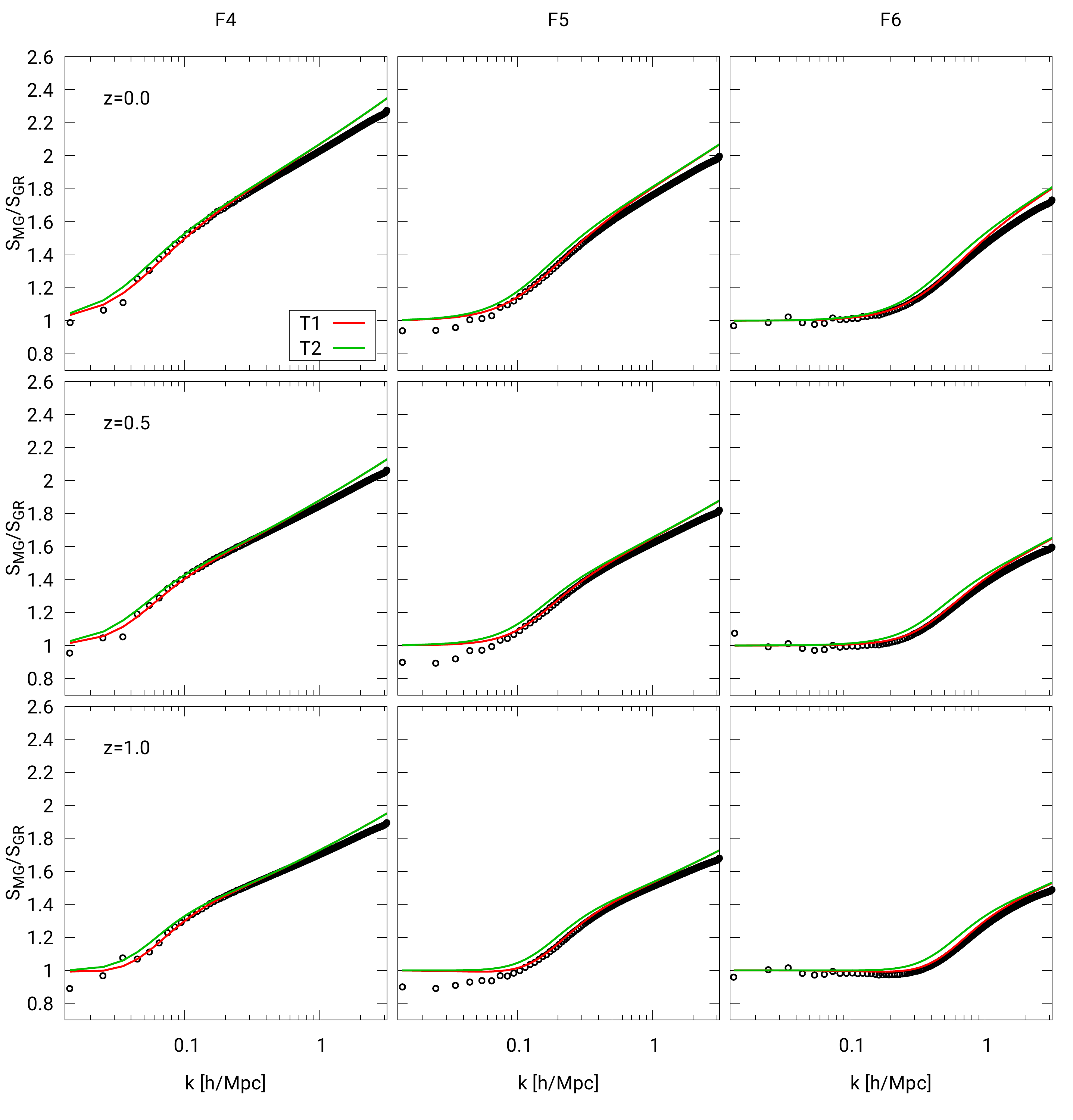}
  \caption{Computation of the full source term (black dots), compared to the
    triangle configurations T1 (orthogonal: $k_1=k_2$, $\theta=90^{\circ}$,
    red line) and T2 ($k_1= k_2$, $\theta=80^{\circ}$, green line), for
    different redshifts and different values of the $\mathrm{f_{R0}}$
    parameter. Top row is z=0, middle row z=0.5 and bottom row is for z=1. The
    left column is $\mathrm{f_{R0}}=-10^{-4}$, middle column
    $\mathrm{f_{R0}}=-10^{-5}$ and right column
    $\mathrm{f_{R0}}=-10^{-6}$. For each redshift and each value of
    $\mathrm{f_{R0}}$ we compute the source term for a box with $700^3$
    particles and $\mathrm{L} = 700 \; {\rm Mpc} \; h^{-1}$.}
  \label{fig:more-models}
\end{figure*}

To understand the generality of this result, we perform the same computation
for three different redshifts ($z=0$, $z=0.5$ and $z=1$) and three different
values of the $f_{R0}$ parameter ($f_{R0}=-10^{-4}$, F4; $f_{R0}=-10^{-5}$,
F5; $f_{R0}=-10^{-6}$, F6). The result is shown in
fig. \ref{fig:more-models}. The black dots represent the result of the ratio
of source terms $S_{MG}/S_{GR}$, while the solid lines represent the two best
triangles found for the F4, $z=0$ case: T1 in red and T2 in green.  We note
that, when considering different redshifts and values of $f_{R0}$, the T1
configuration approximates better the full source term, therefore we adopt it
to compute the approximate $D_2(k,a)$ in the comparison to full N--body
simulations.
%
%
\section{Test against N--body simulations}
\label{sec:test}
To test how well our approximation for second order displacements does at
reconstructing the positions of dark matter halos, we use a suite of N--body
simulations run with $f(R)$ gravity \citep{giocoli2018}, the
DUSTGRAIN--pathfinder simulations. These simulations are performed with the
{\sc mg--gadget} code \citep{puchwein2013} and consist of $768^3$ particles of
mass $8.1 \times 10^{10} M_{\odot}$ in a $750 \; \mathrm{Mpc} \; h^{-1}$ side
box. The adopted cosmology comes from Planck 2015 \citep{planck2016}:
$\Omega_m = 0.31345$, $\Omega_b = 0.0481$, $\Omega_{\Lambda} = 0.68655$, $H_0
= 67.31 \mathrm{km \; s^{-1} \; Mpc^{-1}}$, $A_s = 2.199 \times 10^{-9}$, $n_s
= 0.9658$. The MG model is Hu--Sawicki $f(R)$ with n=1, and three different
values of $f_{R0}= -10^{-4}$ (F4), $-10^{-5}$ (F5), $-10^{-6}$ (F6). For our
tests, we use the simulation with $f_{R0} = -10^{-4}$ to maximize deviations
from GR, and we compare the halo power spectrum we derive to the one measured
in the simulations. A reference $\Lambda$CDM simulation is also available.
Halos are found by running a standard friends-of-friends halo finder on the
simulation snapshots, using a linking length of $0.2$ times the
inter--particle distance.

\begin{figure}
\centering
  \includegraphics[width=\columnwidth]{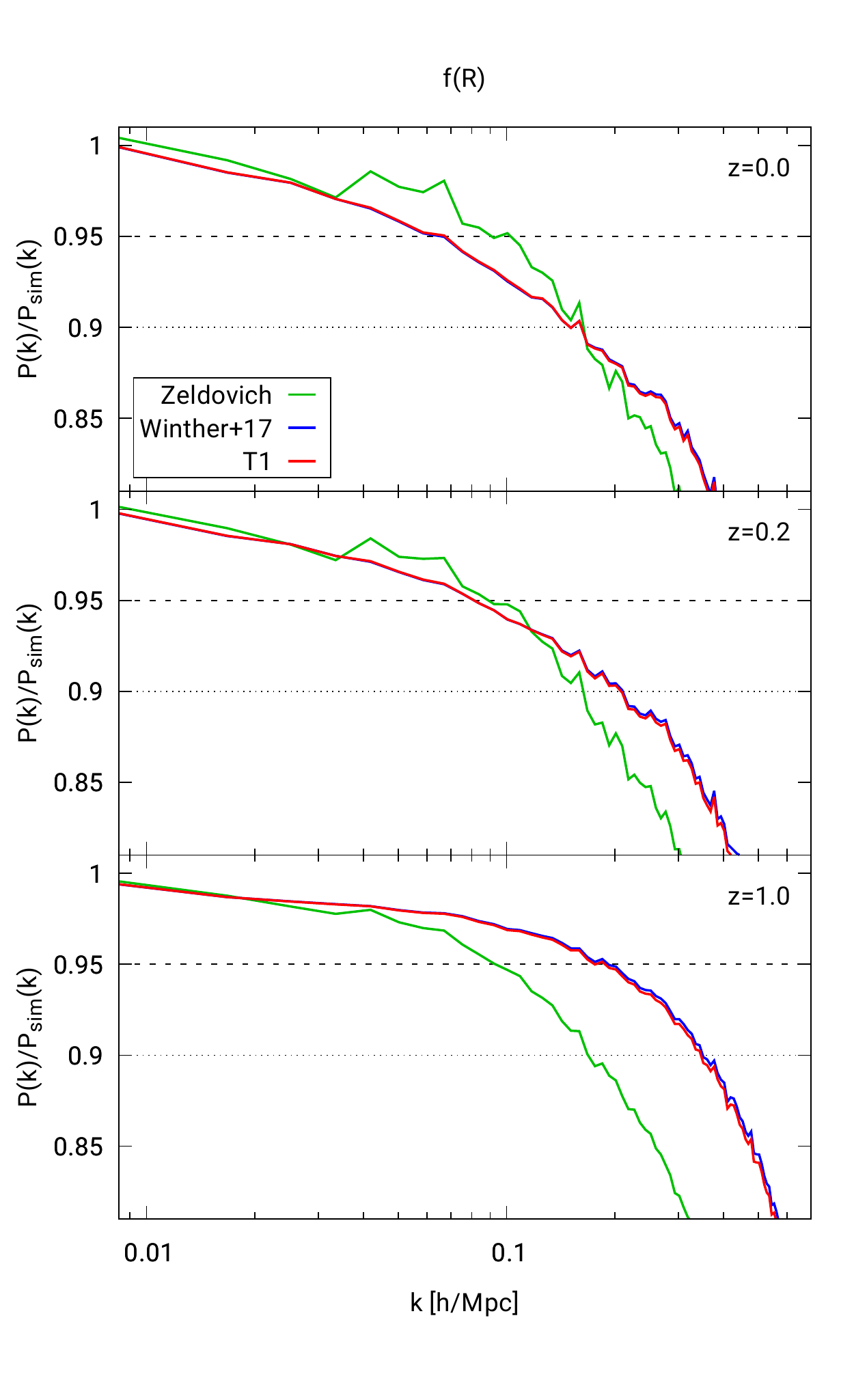}
  \caption{Ratio of the halo power spectrum evaluated with different
    approximations to the one measured from simulations: in green is the
    Zel'dovich approximation, the red line is the T1 triangle with $k_1 =
    k_2$, $\theta = 90^o$. In blue we also plot the result obtained when
    adopting the approximation proposed in \citet{winther2017}. The dashed and
    dotted black lines mark respectively $5\%$ and $10\%$ deviation.}
  \label{fig:sim-comparison}
\end{figure}

\begin{figure}
\centering
  \includegraphics[width=\columnwidth]{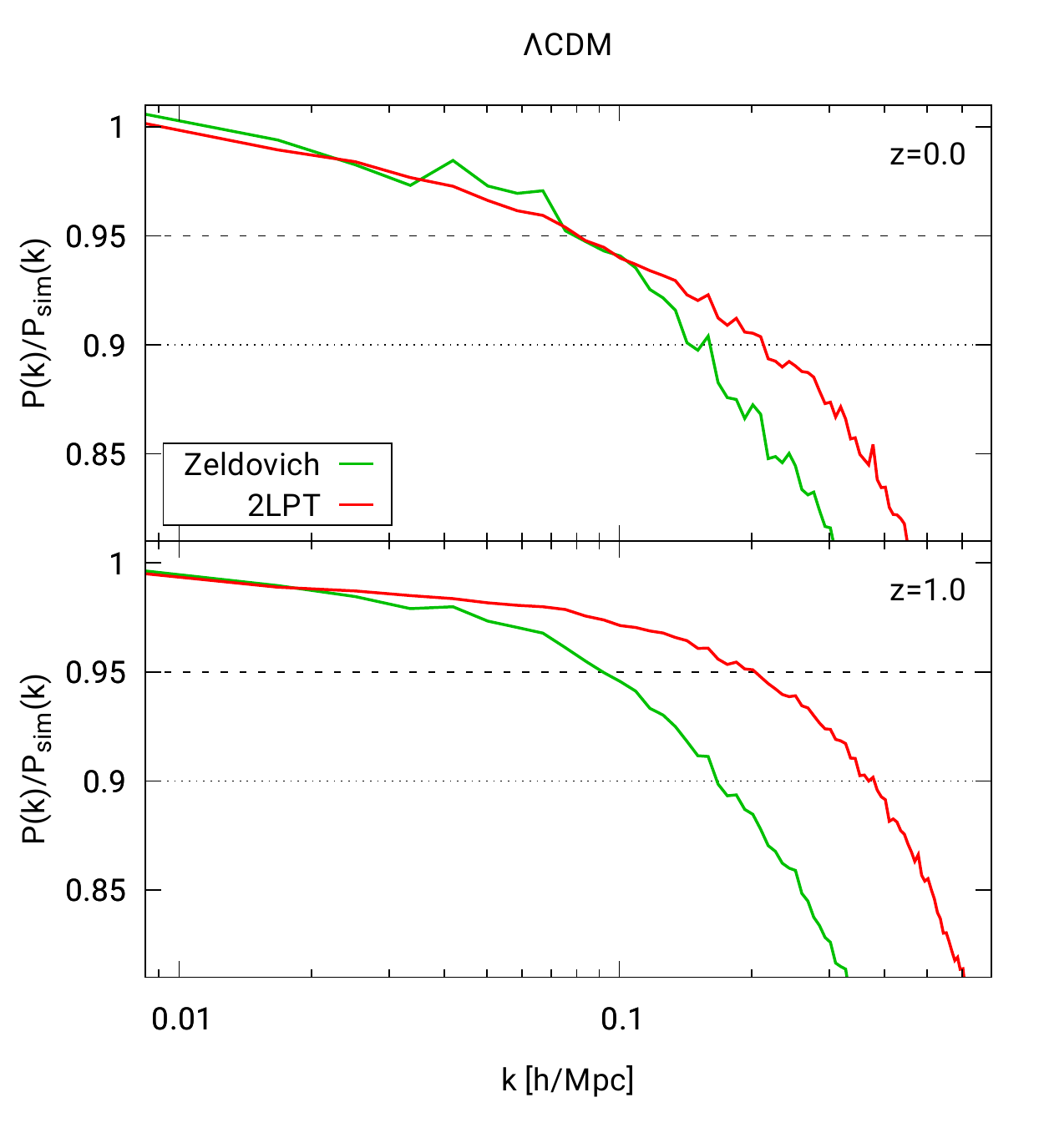}
  \caption{Ratio of the halo power spectrum for $\Lambda$CDM at $z=0$ (top
    panel) and $z=1$ (bottom panel) with respect to the simulation. The
    particles are displaced with the Zel'dovich approximation (first order
    LPT, green line) or second order LPT (red line). The dashed and dotted black
    lines mark respectively $5\%$ and $10\%$ deviation.}
  \label{fig:displ-lcdm}
\end{figure}

Our goal is to assess the performance of our approximation for 2LPT in the
context of modified gravity models.  For this purpose, we conduct an analysis
similar to the one carried out in \citet{munari2017-1}: we set up our code
using the same ICs of the N--body simuation, distributing particles on a
regular grid.  Particles in the same Lagrangian positions are labelled with
the same IDs as in the N--body simulation.  We displace particles using our
approximation for second order LPT and group them in halos using the same
membership of the simulation.  Finally, we construct the halo catalog,
computing the position of each halo by averaging over the particles that
belong to it.  From our reconstructed catalog we evaluate the halo power
spectrum, using the method described in \citet{sefusatti2016}, both for our
catalog and the simulation's one. The result is shown in
fig. \ref{fig:sim-comparison} for three different redshifts: $z=0, 0.2, 1.0$.
Here we plot the ratio of the halo power spectrum obtained when displacing
particles with our approximation to the one measured from simulations.  We
show results for the Zel'dovich approximation (green lines) and for 2LPT
approximated with the T1 triangle configuration (red lines), as well as the
approximation proposed by \citet{winther2017} (blue lines).  The same
quantities are computed for a $\Lambda$CDM simulation and plotted in
fig. \ref{fig:displ-lcdm} at redshift $z=0$ (top panel) and $z=1$ (bottom
panel); here the green line is again the Zel'dovich approximation, while the
red line is 2LPT.

Since the fifth force introduced by the gravity modification enhances the
clustering of matter, the value of $\sigma_8$ at $z=0$ is larger for the $f(R)$
simulation than the $\Lambda$CDM one.  In a sense, at a given redshift a
Universe with MG is \textit{more non--linear} with respect to one where
gravity is described by General Relativity.  Given that the perturbative
approach breaks down as the field becomes non--linear, a fair comparison
between MG and $\Lambda$CDM should be performed between snapshots with the
same level of non-linearity.  To assess the performance of our method with
$f(R)$ gravity with respect to $\Lambda$CDM we choose then two snapshots with
the same value of $\sigma_8$, and compare the halo power spectrum obtained for
$\Lambda$CDM at redshift $z=0$ (top panel of fig. \ref{fig:displ-lcdm}) to the
$f(R)$ one at $z=0.2$ (middle panel of fig. \ref{fig:sim-comparison}).

\begin{figure}
\centering
  \includegraphics[width=\columnwidth]{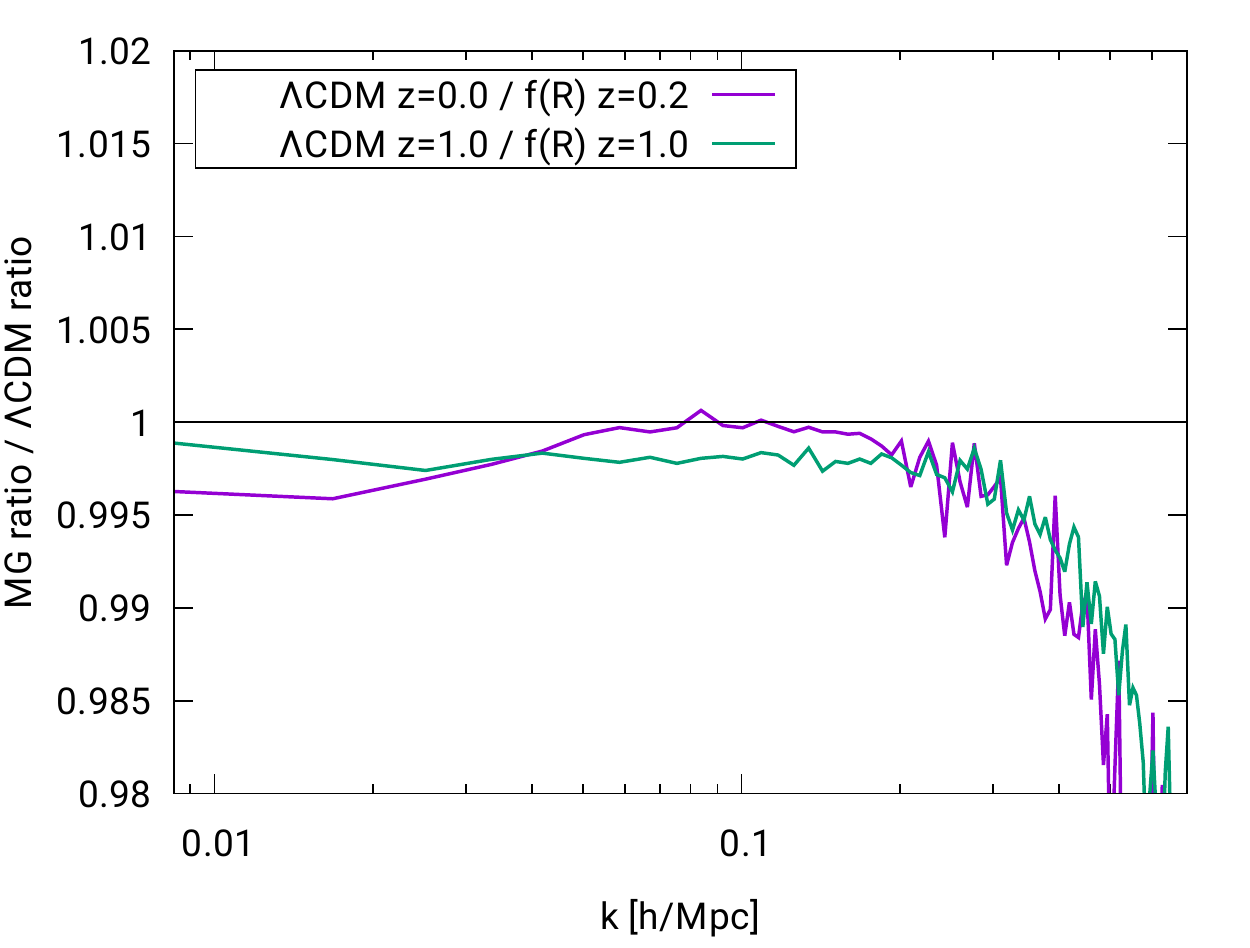}
  \caption{Ratio of the orange lines of fig. \ref{fig:sim-comparison} to the
    blue lines of fig. \ref{fig:displ-lcdm} for $z=1$ (green line) and $z=0.2$
    for MG vs $z=0$ for $\Lambda$CDM (purple line).}
  \label{fig:performance}
\end{figure}

In both cases, the second order approximation allows to reproduce the halo
power spectrum within $10\%$ up to $k \simeq 0.4 \; h \; \mathrm{Mpc}^{-1}$ at
$z=1$ and $k \simeq 0.2 \; h \; \mathrm{Mpc}^{-1}$ at $z=0.2$ for $f(R)$.
This result is very close to the one obtained for 2LPT with $\Lambda$CDM; to
better quantify the performance of 2LPT with modified gravity, we plot in
fig. \ref{fig:performance} the ratio $(P_{MG}(k)/P_{sim,MG}(k)) / (P_{\Lambda
  CDM}(k)/P_{sim,\Lambda CDM})$: the deviation between the two is within $1\%$
up to scales $k \simeq 0.4 \; h \; \mathrm{Mpc}^{-1}$.  Moreover, we can see
from fig. \ref{fig:sim-comparison} that the two approximations we considered
(T1 and the one proposed in \citealt{winther2017}) yield very similar results
in terms of the halo power spectrum, even though they showed a few percent
difference with respect to the full source term.

\begin{figure}
\centering
  \includegraphics[width=\columnwidth]{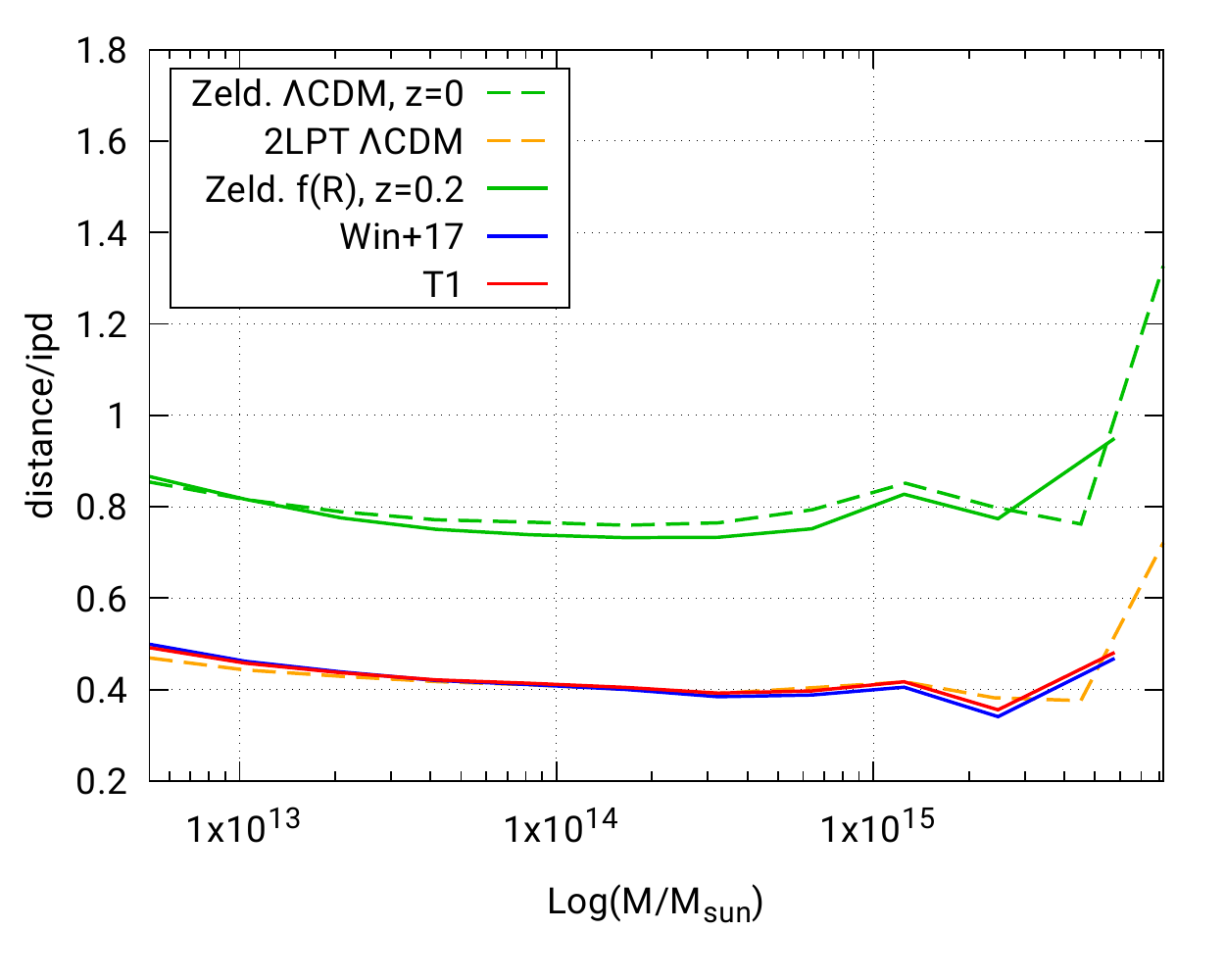}
  \caption{Distance between the halo center as measured in the simulations to
    the one measured after halo reconstruction with the method described in
    section \ref{sec:test} as a function of the halo mass, in units of
    inter--particle distance (\textit{ipd}, $0.977 \; \mathrm{Mpc} \; h^{-1}$).
    Dashed lines represent the median of the halo distance for the
    $\Lambda$CDM simulation at $z=0$, while solid lines represent the same
    quantity for the $f(R)$ simulation at redshift $z=0.2$.  Green lines refer
    to particles displaced with first order LPT (Zel'dovich approximation),
    while the orange, blue and red lines represent respectively the
    $\Lambda$CDM 2LPT, the approximation used in \citet{winther2017} for
    $f(R)$ 2LPT and the triangle configuration labelled as T1 in the previous
    plots.}
  \label{fig:halo-pos-z0}
\end{figure}

\begin{figure}
\centering
  \includegraphics[width=\columnwidth]{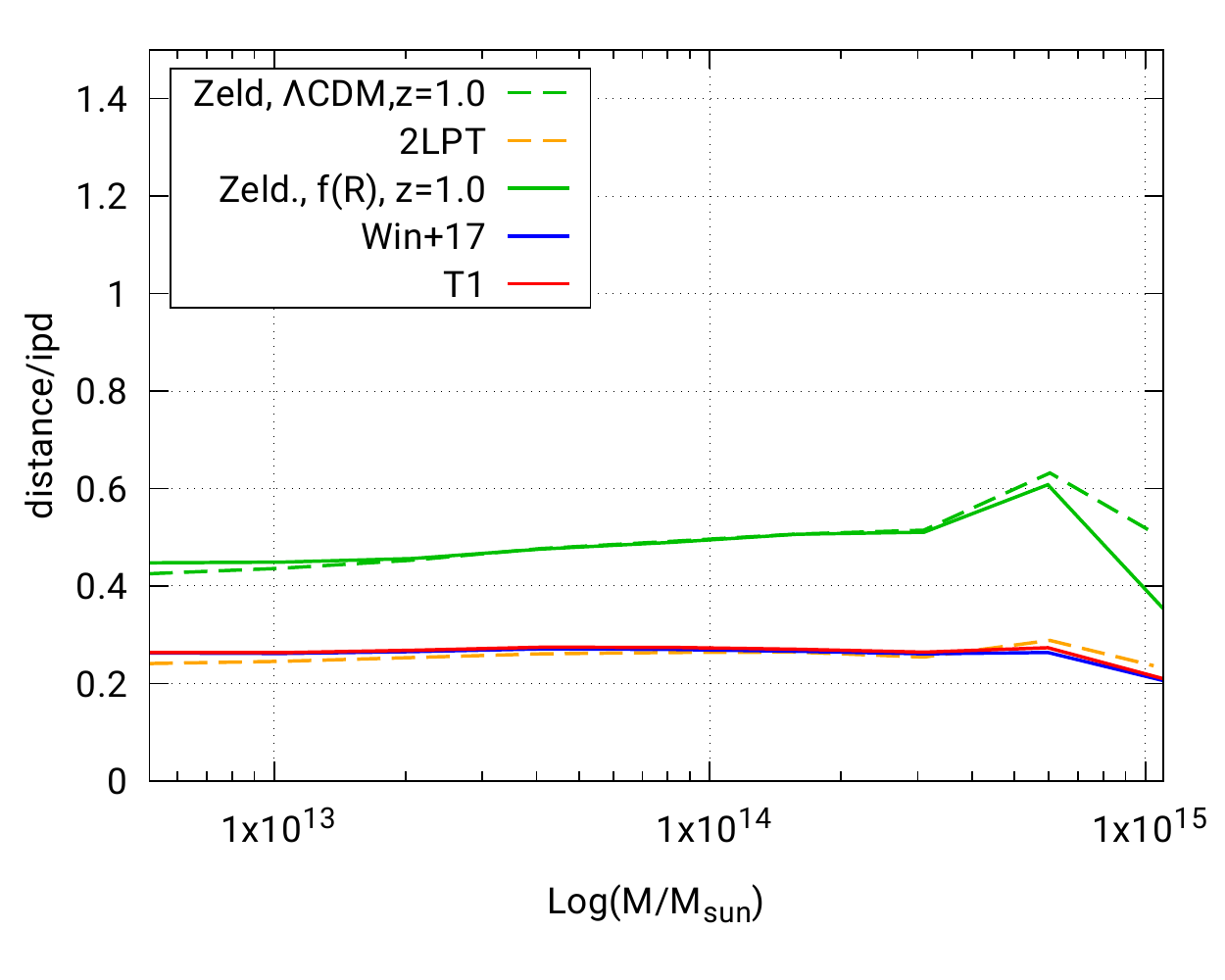}
  \caption{Same as fig. \ref{fig:halo-pos-z0} but at redshift $z=1$.}
  \label{fig:halo-pos-z1}
\end{figure}

We also perform a test to check the accuracy with which we reproduce the halo
centers from particles displaced with our approximation, with respect to the
simulation catalogs.  The result is shown in fig. \ref{fig:halo-pos-z0} and
fig. \ref{fig:halo-pos-z1}, both for the first--order Zel'dovich approximation
(green lines) and 2LPT (red and blue lines, same color--coding as in
fig. \ref{fig:sim-comparison}, with the case of $\Lambda$CDM 2LPT plotted in
orange).  Here we plot the distance between the halo--centers of the
simulation and the ones in our catalog, normalized to the inter--particle
distance (corresponding to $\sim 0.78 \; \mathrm{Mpc} \; h^{-1}$), as a
function of the halo mass.  To assess the performance of our 2LPT+MG approach,
we compute halo distances also for the $\Lambda$CDM scenario (dashed lines in
fig. \ref{fig:halo-pos-z0} and \ref{fig:halo-pos-z1}).  As before, in order to
do a fair comparison between the perturbative approaches in the two gravity
models with the same level of non--linearity, we compare the $\Lambda$CDM one
at $z=0$ to the MG one at $z=0.2$ (fig.\ref{fig:halo-pos-z0}).  It can be seen
that, even though there is on average an error of $\sim 0.8$ times the
inter--particle distance (green lines) for the first order, and $\sim 0.4$
times the inter--particle distance for the second order, the performance is the
same as the one shown by 2LPT+$\Lambda$CDM.  Moreover, the error on the halo
position is roughly independent from the halo mass.  In
fig. \ref{fig:halo-pos-z1} we perform the same test but at redshift $z=1$; as
expected, the LPT halo centers are a better match to the simulation ones', and
the performance for the MG model is again similar to the one obtained for the
standard scenario.
%
%
\section{Testing a fit for $D_2$}
\label{sec:test-lcdm-fit}
We test here if the technique used in \cite{rizzo2017} for neutrinos gives
acceptable results also in the case of $f(R)$ gravity.  Massive neutrinos'
free streaming imprints a scale--dependence to the growth of structures.  The
approach adopted in \cite{rizzo2017} to extend the {\sc pinocchio} code to
massive neutrino cosmologies is based on computing $D_1^2(k,a)$ as the ratio
of the linear power spectrum evaluated at a generic $a$, over the same
quantity calculated at a fixed time $\bar{a}$, where the latter is taken as
the scale factor ensuring $D_1(k,\bar{a}) = 1$.  Linear power spectra are
computed with the {\sc camb} code \citep{lewis2002}.  The second--order growth
factor $D_2(t,k)$ is then computed by adopting the well--known fit, shown to
be valid for a $\Lambda$CDM Universe with standard GR \citep{bouchet1995}:
\begin{equation}
\label{eq:lcdm-fit}
D_2(k,a) = - \frac{3}{7} D_1^2(k,a) \Omega_m(a)^{-1/143} \; .
\end{equation}
We adopt the same approach, to assess if it can be employed in the case of
$f(R)$ gravity.  To this purpose, we used {\sc eftcamb} \citep{hu2015} to
produce linear power spectra (computed for the same Hu--Sawicki $f(R)$ model
discussed before) for a set of redshifts, and then input these power spectra to
the code to compute the linear and second-order growth rate as described
above.

\begin{figure}
  \centering
  \includegraphics[width=\columnwidth]{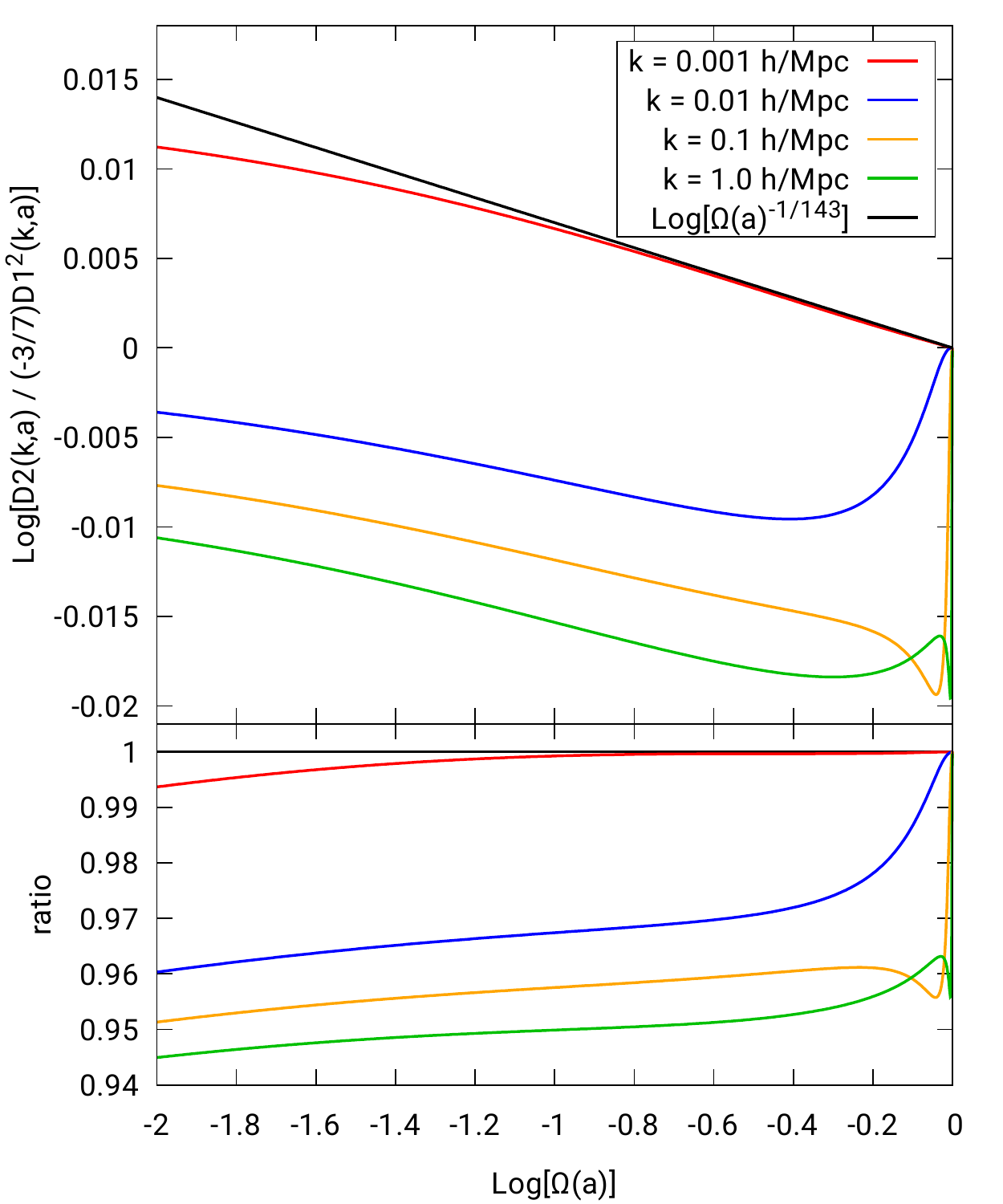}
  \caption{\textit{Top panel}: Ratio of the second order scale--dependent
    growth factor $D_2(k,a)$ to $(-3/7)D_1^2(k,a)$, as a function of
    $\Omega_m(a)$. The black line is the fit of \citet{bouchet1995},
    $\Omega_m(a)^{-1/143}$, while the red, blue, orange and green lines show
    $D_2(k,a)$ for different values of $k$, respectively $0.001$, $0.01$,
    $0.1$ and $1$ $h/ \mathrm{Mpc}$. The modified gravity model chosen is
    $n=1$ Hu--Sawicki with $f_{R0}=-10^{-4}$. \textit{Bottom panel}: Ratio of
    $D_2(k,a) / (-3/7)D_1^2(k,a)$ to $\Omega_m(a)^{-1/143}$. For small values
    of $k$ (red line) the fit of \citet{bouchet1995} is still valid, as
    expected, however, already for $k=0.01 h / \mathrm{Mpc}$, there is a
    deviation of $\sim 3-4 \%$.}
  \label{fig:d2fit}
\end{figure}

\begin{figure}
\centering
  \includegraphics[width=\columnwidth]{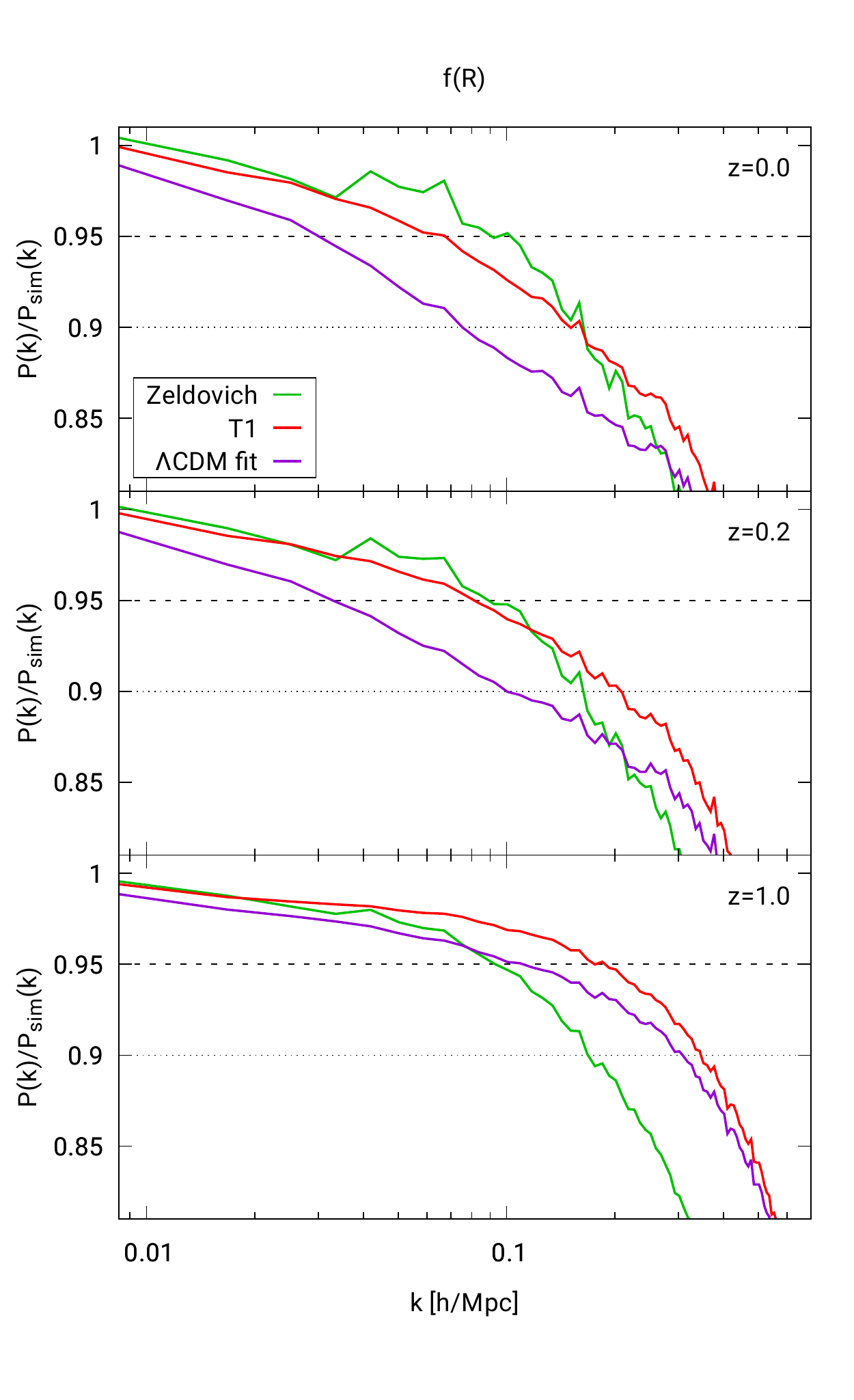}
  \caption{Comparison between the result obtained for the halo power spectrum
    when using the fit valid for $\Lambda$CDM (eq. \ref{eq:lcdm-fit}, shown in
    purple) to compute the second order growth rate and the same quantity
    obtained when $D_2$ is computed by solving the differential equation for
    the T1 triangle configuration (red line).  In green we also show the
    result obtained for the first order LPT, with $D_1(k,a)$ obtained by means
    of ratios of linear power spectra computed with {\sc eftcamb}.}
  \label{fig:2lpt-lcdm-fit}
\end{figure}

In fig. \ref{fig:d2fit} we compare the second order growth rate obtained from
eq. \ref{eq:lcdm-fit} to the one obtained by solving the second order
differential equation for the triangle T1 ($k_1 = k_2$, $\theta =
90^{\circ}$).  In the top panel of fig. \ref{fig:d2fit} we plot the ratio
between $D_2(k,a)$ and $-3 D_1^2(k,a) / 7$ as a function of $\Omega_m(a)$. The
black line represents the best fit obtained by \citet{bouchet1995} for a
$\Lambda$CDM Universe ($\Omega_m(a)^{-1/143}$), while the red, blue, orange
and green lines show the ratio $D_2/(-3 D_1^2/7)$ in the case of Hu--Sawicki
$f(R)$ with $f_{R0}=-10^{-4}$, for increasing value of the wavenumber $k$ as
specified in the legend. The bottom panel shows the ratio of the lines of the
top panel to $\Omega_m(a)^{-1/143}$.  It can be seen that, in the case of
scale--dependent growth induced by modified gravity, eq. \ref{eq:lcdm-fit}
does not provide a good descripion for $D_2$. In particular, even though the
approximation is still accurate for the largest scales ($10^{-3} \, h \,
\mathrm{Mpc}^{-1}$, red line), where we do not expect significant effects of
MG on the growth rates, for smaller scales (and already at $k = 10^{-2} \, h
\, \mathrm{Mpc}^{-1}$, blue line), the growth rate deviates for more than
$\sim 3-4 \%$ from the fit, and the deviation gets stronger as we go to
smaller scales. This is due to the fact that the scale dependence of
$D_2(k,a)$ is not accurately modelled by $D_1^2(k,a)$. To properly treat
mildly non--linear scales we cannot use the fit of eq. \ref{eq:lcdm-fit}, and
must therefore resort to the method described in the previous sections.  The
result of using this approximation to compute $D_2$ is shown in
fig. \ref{fig:2lpt-lcdm-fit}: here we plot again the ratio of the halo power
spectrum obtained with 2LPT to the N--body simulation one, and compare it to
the one computed with the T1 triangle configuration.  It is clear that the
results obtained with the T1 triangle (red lines) are a better match to the
simulation's halo $P(k)$ than the one obtained when using
eq. \ref{eq:lcdm-fit} (purple line).  In particular, when adopting
eq. \ref{eq:lcdm-fit} to compute second order displacements at $z=1$, the
resulting halo power spectrum does not show any improvements with respect to
the linear approsimation for scales $0.04\ h\ {\rm Mpc}^{-1} \leq k \leq
0.1\ h\ {\rm Mpc}^{-1}$ (bottom panel of fig. \ref{fig:2lpt-lcdm-fit}).
%
%
\section{Conclusions}
\label{sec:conclusion}
Future generations of galaxy redshift surveys will allow to measure the
clustering of matter with a high degree of accuracy, allowing in principle to
disentangle between different gravity theories.  In order to test alternatives
to General Relativity, a proper treatment of non--linear and quasi non--linear
scales is required, since these are the scales where possible deviations from
GR can be found.  An adequate description of quasi non--linear scales can be
achieved via N--body simulations or, alternatively, with approximate methods.
The latter allow, with some compromises on the accuracy, to generate the large
sets of simulated catalogs needed to accurately constrain the cosmological
parameters, a task that cannot be pursued with the computationally expensive
N--body simulations.  This work fits in the framework of extending these
approximated methods to modified gravity theories, focusing in particular on
the computation of second--order Lagrangian displacements.

We presented a new computation of second-order LPT that is valid for a class
of modified gravity theories, and specialized it to the case of Hu--Sawicki
$f(R)$ theory, testing its performances against N--body simulations.  In MG
theories the various expansion terms of LPT are typically not separable as
products of time--dependent and space--dependent functions, and the equation
for the second--order Fourier--space Lagrangian potential
$\phi^{(2)}(\vec{k},a)$ can be written as an integral over two more vectors
$\vec{k}_1$ and $\vec{k}_2$, that are constrained to form a triangle with
$\vec{k}$.  For the case in which the coefficients $M_k$
(eq. \ref{eq:mg-pot-exp}) of the Taylor expansion of the fluctuations in the
Ricci scalar $\delta R$ are not scale-dependent, the differential equation for
the 2LPT displacement potential can be written in terms of direct and inverse
Fourier transforms. This allows to treat it with a numerical approach.

Using an initial density field sampled in cubic boxes of varying size and
number of grid points, we numerically characterized the source term of the
2LPT potential (normalized by its GR counterpart) by computing its average and
standard deviation as a function of $k$.  We then considered different
triangle configurations to find the second order growth factor $D_2(k, k_1,
k_2, a)$ that best reproduces the average of the source term, and used it to
achieve an effective separation of the 2LPT displacement field into a space
part, that does not depend on time and is equal to that used in GR, and a
$k$-dependent second--order growth rate $D_2(k,a)$.  The latter can be
computed by numerically integrating a set of Ordinary Differential Equations,
one for each $k$ value.  The scatter in the numerical solution around the
average source term gives a measure of the accuracy of this approximation, and
is found to be moderate at the scales where 2LPT is relevant.  We also tested
the approximations we chose for $D_2(k,a)$ at different redshifts and for
different values of the $f_{R0}$ parameter, and found that the chosen triangle
configurations can be safely adopted.

We implemented the solution for both differential equations for $D_2(k,a)$ in
our code to compute Lagrangian displacements, and followed the approach
discussed in \cite{munari2017-1} to test the accuracy level to which we can
reproduce halo positions with respect to an N--body simulation.  We produced a
second--order displacement field, and compared with the results of a
simulation run with MG--Gadget \citep{giocoli2018} and Hu--Sawicki $f(R)$
gravity (with a large value of $f_{R0}=-10^{-4}$, to maximise the effect of
modified gravity).  The halos in the simulation were identified by using a
standard friends-of-friends halo finder algorithm.  To construct our halo
catalog, we used the same particle assignment of the simulation to group
particles displaced with 2LPT, then we re-computed each halo center of mass as
the average over all particles that belong to it.  Using these halo
displacements we computed the halo power spectrum and compared it with that
measured from the N--body halo catalog.  As demonstrated by
\cite{munari2017-1} in the context of $\Lambda$CDM, this procedure allows to
test how an approximate method like 2LPT can recover the clustering of halos
without being required to solve the much harder problem of identifying halos
themselves.  We find that both chosen triangle configurations, together with
the one previously proposed by \citet{winther2017}, perform well in terms of
the halo power spectrum, allowing to reconstruct it within $\sim 10\%$ at
mildly non linear scales ($k \simeq 0.2 - 0.4 \; h \; \mathrm{Mpc}^{-1}$).
This performance is the same (within $1\%$) as the one shown by 2LPT in a
standard, $\Lambda$CDM Universe with General Relativity, as highlighted in
fig. \ref{fig:performance}, meaning that the loss of power in our
reconstructed halo $P(k)$ with respect to the N--body one is mostly due to the
failure of the perturbative approach as the displacement field becomes
non--linear.  We conclude that LPT can be safely used to displace particles
even in presence of modified gravity.

The method we employ to construct the halos, by matching the particle
memberships to the simulation ones, means that we can perform an
object--by--oject analysis.  We therefore verify how good our approximation
for the halo displacements is at recovering the halo positions with respect to
the simulation.  The result is again consistent with the one obtained in a
$\Lambda$CDM scenario.

Throughout this work, we focused on a particular class of modified gravity
models, Hu--Sawicki $f(R)$.  The method we propose is however quite general,
and can be extended to other MG theories: once the functional form for the
$\mu(k,a)$ function (that parametrizes the Fourier--space Poisson equation)
and the $M_k$ coefficients are known, the procedure we propose can be employed
to find a proper approximation for $D_2$. 
If the $M_k$ coefficients are scale--dependent the method can still be
applied, provided that the $S_3$ (scalar field self--interaction) term of
eq. \ref{eq:s3} can be written in terms of Fourier transforms.  This requires
identifying the proper operators in configuration space that correspond to the
$M_k$ coefficients in Fourier space.
We stress that the procedure must be done only once for each gravity theory,
and does not require the use of N--body simulations.

This allows to produce large sets of approximated simulations for different
gravity models, a task that plays a crucial role in the computation of the
covariance matrices needed to constrain cosmological parameters.  We
implemented MG scale--dependent growth in the {\sc pinocchio} code as an
optional functionality, making it able to generate 2LPT displacements fields
with modified gravity. However, a key part of the algorithm is the one that
groups particles in halos, needed to make the code fully predictive.  In the
standard {\sc pinocchio} code this is done by treating overdensities as
homogeneous ellipsoids, and computing collapse times as the moment of first
orbit crossing. This part of the algorithm still needs to be adapted to
modified gravity, so that the code can generate halo catalogs
independently. This will involve formulating ellipsoidal collapse with
modified gravity, and is the focus of a future work.
%
%
\section*{Acknowledgements}
The authors warmly acknowledge many discussions with the {\sc EFTCAMB} team,
in particular with Alessandra Silvestri, Bin Hu, Marco Raveri and Jorgos
Papadomanolakis. This paper has benefited from the stimulating environment of
the Euclid Consortium.  C.M. and P.M. acknowledge support from PRIN MIUR 2015
{\em Cosmology and Fundamental Physics: illuminating the Dark Universe with
  Euclid} and from a {\em Fondo Ricerca di Ateneo} of the Trieste University.
P.M. has been supported by INFN InDark research project.
TheDUSTGRAIN-pathfinder simulations employed in this work have been performed
on the Marconi supercomputing machine at Cineca thanks to the PRACE project
SIMCODE1 (grant nr. 2016153604) and on the computing facilities of the
Computational Center for Particle and Astrophysics (C2PAP) and of the Leibniz
Supercomputer Center (LRZ) under the project ID pr94ji.
%
%

\bibliographystyle{mnras} \bibliography{refs}

\begin{thebibliography}{}
\makeatletter
\relax
\def\mn@urlcharsother{\let\do\@makeother \do\$\do\&\do\#\do\^\do\_\do\%\do\~}
\def\mn@doi{\begingroup\mn@urlcharsother \@ifnextchar [ {\mn@doi@}
  {\mn@doi@[]}}
\def\mn@doi@[#1]#2{\def\@tempa{#1}\ifx\@tempa\@empty \href
  {http://dx.doi.org/#2} {doi:#2}\else \href {http://dx.doi.org/#2} {#1}\fi
  \endgroup}
\def\mn@eprint#1#2{\mn@eprint@#1:#2::\@nil}
\def\mn@eprint@arXiv#1{\href {http://arxiv.org/abs/#1} {{\tt arXiv:#1}}}
\def\mn@eprint@dblp#1{\href {http://dblp.uni-trier.de/rec/bibtex/#1.xml}
  {dblp:#1}}
\def\mn@eprint@#1:#2:#3:#4\@nil{\def\@tempa {#1}\def\@tempb {#2}\def\@tempc
  {#3}\ifx \@tempc \@empty \let \@tempc \@tempb \let \@tempb \@tempa \fi \ifx
  \@tempb \@empty \def\@tempb {arXiv}\fi \@ifundefined
  {mn@eprint@\@tempb}{\@tempb:\@tempc}{\expandafter \expandafter \csname
  mn@eprint@\@tempb\endcsname \expandafter{\@tempc}}}

\bibitem[\protect\citeauthoryear{{Amendola} et~al.,}{{Amendola}
  et~al.}{2018}]{amendola2018}
{Amendola} L.,  et~al., 2018, \mn@doi [Living Reviews in Relativity]
  {10.1007/s41114-017-0010-3}, \href
  {https://ui.adsabs.harvard.edu/abs/2018LRR....21....2A} {21, 2}

\bibitem[\protect\citeauthoryear{{Avila}, {Murray}, {Knebe}, {Power},
  {Robotham}  \& {Garcia-Bellido}}{{Avila} et~al.}{2015}]{avila2015}
{Avila} S.,  {Murray} S.~G.,  {Knebe} A.,  {Power} C.,  {Robotham} A.~S.~G.,
  {Garcia-Bellido} J.,  2015, \mn@doi [\mnras] {10.1093/mnras/stv711}, \href
  {http://adsabs.harvard.edu/abs/2015MNRAS.450.1856A} {450, 1856}

\bibitem[\protect\citeauthoryear{{Aviles} \& {Cervantes-Cota}}{{Aviles} \&
  {Cervantes-Cota}}{2017}]{aviles2017}
{Aviles} A.,  {Cervantes-Cota} J.~L.,  2017, \mn@doi [\prd]
  {10.1103/PhysRevD.96.123526}, \href
  {http://adsabs.harvard.edu/abs/2017PhRvD..96l3526A} {96, 123526}

\bibitem[\protect\citeauthoryear{{Baldi}, {Villaescusa-Navarro}, {Viel},
  {Puchwein}, {Springel}  \& {Moscardini}}{{Baldi} et~al.}{2014}]{baldi2014}
{Baldi} M.,  {Villaescusa-Navarro} F.,  {Viel} M.,  {Puchwein} E.,  {Springel}
  V.,   {Moscardini} L.,  2014, \mn@doi [\mnras] {10.1093/mnras/stu259}, \href
  {https://ui.adsabs.harvard.edu/abs/2014MNRAS.440...75B} {440, 75}

\bibitem[\protect\citeauthoryear{{Blot} et~al.,}{{Blot}
  et~al.}{2019}]{blot2019}
{Blot} L.,  et~al., 2019, \mn@doi [\mnras] {10.1093/mnras/stz507}, \href
  {http://adsabs.harvard.edu/abs/2019MNRAS.485.2806B} {485, 2806}

\bibitem[\protect\citeauthoryear{{Bond} \& {Myers}}{{Bond} \&
  {Myers}}{1996}]{bond1996}
{Bond} J.~R.,  {Myers} S.~T.,  1996, \mn@doi [\apj] {10.1086/192267}, \href
  {http://adsabs.harvard.edu/abs/1996ApJS..103....1B} {103, 1}

\bibitem[\protect\citeauthoryear{{Bouchet}}{{Bouchet}}{1996}]{bouchet1996}
{Bouchet} F.~R.,  1996, in {Bonometto} S.,  {Primack} J.~R.,   {Provenzale} A.,
   eds, Dark Matter in the Universe. p.~565 (\mn@eprint {} {astro-ph/9603013})

\bibitem[\protect\citeauthoryear{{Bouchet}, {Colombi}, {Hivon}  \&
  {Juszkiewicz}}{{Bouchet} et~al.}{1995}]{bouchet1995}
{Bouchet} F.~R.,  {Colombi} S.,  {Hivon} E.,   {Juszkiewicz} R.,  1995, \aap,
  \href {https://ui.adsabs.harvard.edu/abs/1995A&A...296..575B} {296, 575}

\bibitem[\protect\citeauthoryear{{Bull} et~al.,}{{Bull}
  et~al.}{2016}]{bull2016}
{Bull} P.,  et~al., 2016, \mn@doi [Physics of the Dark Universe]
  {10.1016/j.dark.2016.02.001}, \href
  {https://ui.adsabs.harvard.edu/abs/2016PDU....12...56B} {12, 56}

\bibitem[\protect\citeauthoryear{{Burgess}}{{Burgess}}{2013}]{burgess2013}
{Burgess} C.~P.,  2013, arXiv e-prints, \href
  {https://ui.adsabs.harvard.edu/abs/2013arXiv1309.4133B} {}

\bibitem[\protect\citeauthoryear{{Castorina}, {Sefusatti}, {Sheth},
  {Villaescusa-Navarro}  \& {Viel}}{{Castorina} et~al.}{2014}]{castorina2014}
{Castorina} E.,  {Sefusatti} E.,  {Sheth} R.~K.,  {Villaescusa-Navarro} F.,
  {Viel} M.,  2014, \mn@doi [Journal of Cosmology and Astro-Particle Physics]
  {10.1088/1475-7516/2014/02/049}, \href
  {https://ui.adsabs.harvard.edu/abs/2014JCAP...02..049C} {2014, 049}

\bibitem[\protect\citeauthoryear{{Colavincenzo} et~al.,}{{Colavincenzo}
  et~al.}{2019}]{colavincenzo2019}
{Colavincenzo} M.,  et~al., 2019, \mn@doi [\mnras] {10.1093/mnras/sty2964},
  \href {http://adsabs.harvard.edu/abs/2019MNRAS.482.4883C} {482, 4883}

\bibitem[\protect\citeauthoryear{{De Felice} \& {Tsujikawa}}{{De Felice} \&
  {Tsujikawa}}{2010}]{defelice2010}
{De Felice} A.,  {Tsujikawa} S.,  2010, \mn@doi [Living Reviews in Relativity]
  {10.12942/lrr-2010-3}, \href
  {http://adsabs.harvard.edu/abs/2010LRR....13....3D} {13, 3}

\bibitem[\protect\citeauthoryear{{Event Horizon Telescope Collaboration}
  et~al.,}{{Event Horizon Telescope Collaboration} et~al.}{2019}]{eht2019}
{Event Horizon Telescope Collaboration} et~al., 2019, \mn@doi [\apjl]
  {10.3847/2041-8213/ab0ec7}, \href
  {http://adsabs.harvard.edu/abs/2019ApJ...875L...1E} {875, L1}

\bibitem[\protect\citeauthoryear{{Giocoli}, {Baldi}  \& {Moscardini}}{{Giocoli}
  et~al.}{2018}]{giocoli2018}
{Giocoli} C.,  {Baldi} M.,   {Moscardini} L.,  2018, \mn@doi [\mnras]
  {10.1093/mnras/sty2465}, \href
  {http://adsabs.harvard.edu/abs/2018MNRAS.481.2813G} {481, 2813}

\bibitem[\protect\citeauthoryear{{Hu} \& {Sawicki}}{{Hu} \&
  {Sawicki}}{2007}]{hu2007}
{Hu} W.,  {Sawicki} I.,  2007, \mn@doi [\prd] {10.1103/PhysRevD.76.064004},
  \href {https://ui.adsabs.harvard.edu/abs/2007PhRvD..76f4004H} {76, 064004}

\bibitem[\protect\citeauthoryear{{Hu}, {Raveri}, {Silvestri}  \&
  {Frusciante}}{{Hu} et~al.}{2015}]{hu2015}
{Hu} B.,  {Raveri} M.,  {Silvestri} A.,   {Frusciante} N.,  2015, \mn@doi
  [\prd] {10.1103/PhysRevD.91.063524}, \href
  {https://ui.adsabs.harvard.edu/abs/2015PhRvD..91f3524H} {91, 063524}

\bibitem[\protect\citeauthoryear{{Ishak}}{{Ishak}}{2019}]{ishak2019}
{Ishak} M.,  2019, \mn@doi [Living Reviews in Relativity]
  {10.1007/s41114-018-0017-4}, \href
  {https://ui.adsabs.harvard.edu/abs/2019LRR....22....1I} {22, 1}

\bibitem[\protect\citeauthoryear{{Izard}, {Crocce}  \& {Fosalba}}{{Izard}
  et~al.}{2016}]{izard2016}
{Izard} A.,  {Crocce} M.,   {Fosalba} P.,  2016, \mn@doi [\mnras]
  {10.1093/mnras/stw797}, \href
  {https://ui.adsabs.harvard.edu/abs/2016MNRAS.459.2327I} {459, 2327}

\bibitem[\protect\citeauthoryear{{Joyce}, {Jain}, {Khoury}  \&
  {Trodden}}{{Joyce} et~al.}{2015}]{joyce2015}
{Joyce} A.,  {Jain} B.,  {Khoury} J.,   {Trodden} M.,  2015, \mn@doi [\physrep]
  {10.1016/j.physrep.2014.12.002}, \href
  {https://ui.adsabs.harvard.edu/abs/2015PhR...568....1J} {568, 1}

\bibitem[\protect\citeauthoryear{{Kitaura}, {Yepes}  \& {Prada}}{{Kitaura}
  et~al.}{2014}]{kitaura2014}
{Kitaura} F.~S.,  {Yepes} G.,   {Prada} F.,  2014, \mn@doi [\mnras]
  {10.1093/mnrasl/slt172}, \href
  {https://ui.adsabs.harvard.edu/abs/2014MNRAS.439L..21K} {439, L21}

\bibitem[\protect\citeauthoryear{{Koda}, {Blake}, {Beutler}, {Kazin}  \&
  {Marin}}{{Koda} et~al.}{2016}]{koda2016}
{Koda} J.,  {Blake} C.,  {Beutler} F.,  {Kazin} E.,   {Marin} F.,  2016,
  \mn@doi [\mnras] {10.1093/mnras/stw763}, \href
  {https://ui.adsabs.harvard.edu/abs/2016MNRAS.459.2118K} {459, 2118}

\bibitem[\protect\citeauthoryear{{Koyama}, {Taruya}  \& {Hiramatsu}}{{Koyama}
  et~al.}{2009}]{koyama2009}
{Koyama} K.,  {Taruya} A.,   {Hiramatsu} T.,  2009, \mn@doi [\prd]
  {10.1103/PhysRevD.79.123512}, \href
  {http://adsabs.harvard.edu/abs/2009PhRvD..79l3512K} {79, 123512}

\bibitem[\protect\citeauthoryear{{LIGO Scientific Collaboration and Virgo
  Collaboration}, Abbott  \& Abbott}{{LIGO Scientific Collaboration and Virgo
  Collaboration} et~al.}{2016}]{abbott2016}
{LIGO Scientific Collaboration and Virgo Collaboration} Abbott B.~P.,  Abbott
  R.,   Abbott T.~D.,  2016, \mn@doi [Phys. Rev. Lett.]
  {10.1103/PhysRevLett.116.061102}, 116, 061102

\bibitem[\protect\citeauthoryear{{LSST Science Collaboration} et~al.,}{{LSST
  Science Collaboration} et~al.}{2009}]{abell2009}
{LSST Science Collaboration} et~al., 2009, arXiv e-prints, \href
  {http://adsabs.harvard.edu/abs/2009arXiv0912.0201L} {}

\bibitem[\protect\citeauthoryear{{Laureijs} et~al.,}{{Laureijs}
  et~al.}{2011}]{laureijs2011}
{Laureijs} R.,  et~al., 2011, preprint

\bibitem[\protect\citeauthoryear{{Levi} et~al.,}{{Levi}
  et~al.}{2013}]{levi2013}
{Levi} M.,  et~al., 2013, arXiv e-prints, \href
  {http://adsabs.harvard.edu/abs/2013arXiv1308.0847L} {}

\bibitem[\protect\citeauthoryear{Lewis \& Bridle}{Lewis \&
  Bridle}{2002}]{lewis2002}
Lewis A.,  Bridle S.,  2002, \mn@doi [\prd] {10.1103/PhysRevD.66.103511}, 66,
  103511

\bibitem[\protect\citeauthoryear{{Lippich} et~al.,}{{Lippich}
  et~al.}{2019}]{lippich2019}
{Lippich} M.,  et~al., 2019, \mn@doi [\mnras] {10.1093/mnras/sty2757}, \href
  {http://adsabs.harvard.edu/abs/2019MNRAS.482.1786L} {482, 1786}

\bibitem[\protect\citeauthoryear{{Martin}}{{Martin}}{2012}]{martin2012}
{Martin} J.,  2012, \mn@doi [Comptes Rendus Physique]
  {10.1016/j.crhy.2012.04.008}, \href
  {https://ui.adsabs.harvard.edu/abs/2012CRPhy..13..566M} {13, 566}

\bibitem[\protect\citeauthoryear{{Monaco}}{{Monaco}}{2016}]{monaco2016}
{Monaco} P.,  2016, \mn@doi [Galaxies] {10.3390/galaxies4040053}, \href
  {http://adsabs.harvard.edu/abs/2016Galax...4...53M} {4, 53}

\bibitem[\protect\citeauthoryear{{Monaco}, {Theuns}  \& {Taffoni}}{{Monaco}
  et~al.}{2002}]{monaco2002-2}
{Monaco} P.,  {Theuns} T.,   {Taffoni} G.,  2002, \mn@doi [\mnras]
  {10.1046/j.1365-8711.2002.05162.x}, \href
  {http://adsabs.harvard.edu/abs/2002MNRAS.331..587M} {331, 587}

\bibitem[\protect\citeauthoryear{{Munari}, {Monaco}, {Koda}, {Kitaura},
  {Sefusatti}  \& {Borgani}}{{Munari} et~al.}{2017a}]{munari2017-1}
{Munari} E.,  {Monaco} P.,  {Koda} J.,  {Kitaura} F.-S.,  {Sefusatti} E.,
  {Borgani} S.,  2017a, \mn@doi [\jcap] {10.1088/1475-7516/2017/07/050}, \href
  {http://adsabs.harvard.edu/abs/2017JCAP...07..050M} {7, 050}

\bibitem[\protect\citeauthoryear{{Munari}, {Monaco}, {Sefusatti}, {Castorina},
  {Mohammad}, {Anselmi}  \& {Borgani}}{{Munari} et~al.}{2017b}]{munari2017-2}
{Munari} E.,  {Monaco} P.,  {Sefusatti} E.,  {Castorina} E.,  {Mohammad} F.~G.,
   {Anselmi} S.,   {Borgani} S.,  2017b, \mn@doi [\mnras]
  {10.1093/mnras/stw3085}, \href
  {http://adsabs.harvard.edu/abs/2017MNRAS.465.4658M} {465, 4658}

\bibitem[\protect\citeauthoryear{{Perlmutter} et~al.,}{{Perlmutter}
  et~al.}{1999}]{perlmutter1999}
{Perlmutter} S.,  et~al., 1999, \apj, 517, 565

\bibitem[\protect\citeauthoryear{{Planck Collaboration} et~al.,}{{Planck
  Collaboration} et~al.}{2016}]{planck2016}
{Planck Collaboration} et~al., 2016, \mn@doi [\aap]
  {10.1051/0004-6361/201525830}, \href
  {http://adsabs.harvard.edu/abs/2016A%26A...594A..13P} {594, A13}

\bibitem[\protect\citeauthoryear{{Puchwein}, {Baldi}  \& {Springel}}{{Puchwein}
  et~al.}{2013}]{puchwein2013}
{Puchwein} E.,  {Baldi} M.,   {Springel} V.,  2013, \mn@doi [\mnras]
  {10.1093/mnras/stt1575}, \href
  {http://adsabs.harvard.edu/abs/2013MNRAS.436..348P} {436, 348}

\bibitem[\protect\citeauthoryear{{Riess} et~al.,}{{Riess}
  et~al.}{1998}]{riess1998}
{Riess} A.~G.,  et~al., 1998, \mn@doi [\aj] {10.1086/300499}, \href
  {http://adsabs.harvard.edu/abs/1998AJ....116.1009R} {116, 1009}

\bibitem[\protect\citeauthoryear{{Rizzo}, {Villaescusa-Navarro}, {Monaco},
  {Munari}, {Borgani}, {Castorina}  \& {Sefusatti}}{{Rizzo}
  et~al.}{2017}]{rizzo2017}
{Rizzo} L.~A.,  {Villaescusa-Navarro} F.,  {Monaco} P.,  {Munari} E.,
  {Borgani} S.,  {Castorina} E.,   {Sefusatti} E.,  2017, \mn@doi [\jcap]
  {10.1088/1475-7516/2017/01/008}, \href
  {http://adsabs.harvard.edu/abs/2017JCAP...01..008R} {1, 008}

\bibitem[\protect\citeauthoryear{{Sefusatti}, {Crocce}, {Scoccimarro}  \&
  {Couchman}}{{Sefusatti} et~al.}{2016}]{sefusatti2016}
{Sefusatti} E.,  {Crocce} M.,  {Scoccimarro} R.,   {Couchman} H.~M.~P.,  2016,
  \mn@doi [\mnras] {10.1093/mnras/stw1229}, \href
  {http://adsabs.harvard.edu/abs/2016MNRAS.460.3624S} {460, 3624}

\bibitem[\protect\citeauthoryear{{Spergel} et~al.,}{{Spergel}
  et~al.}{2013}]{spergel2013}
{Spergel} D.,  et~al., 2013, arXiv e-prints, \href
  {http://adsabs.harvard.edu/abs/2013arXiv1305.5422S} {}

\bibitem[\protect\citeauthoryear{{Stein}, {Alvarez}  \& {Bond}}{{Stein}
  et~al.}{2019}]{stein2019}
{Stein} G.,  {Alvarez} M.~A.,   {Bond} J.~R.,  2019, \mn@doi [\mnras]
  {10.1093/mnras/sty3226}, \href
  {http://adsabs.harvard.edu/abs/2019MNRAS.483.2236S} {483, 2236}

\bibitem[\protect\citeauthoryear{{Tassev}, {Zaldarriaga}  \&
  {Eisenstein}}{{Tassev} et~al.}{2013}]{tassev2013}
{Tassev} S.,  {Zaldarriaga} M.,   {Eisenstein} D.~J.,  2013, \mn@doi [\jcap]
  {10.1088/1475-7516/2013/06/036}, \href
  {http://adsabs.harvard.edu/abs/2013JCAP...06..036T} {6, 036}

\bibitem[\protect\citeauthoryear{{Valogiannis} \& {Bean}}{{Valogiannis} \&
  {Bean}}{2017}]{valogiannis2017}
{Valogiannis} G.,  {Bean} R.,  2017, \mn@doi [\prd]
  {10.1103/PhysRevD.95.103515}, \href
  {http://adsabs.harvard.edu/abs/2017PhRvD..95j3515V} {95, 103515}

\bibitem[\protect\citeauthoryear{{Weinberg}}{{Weinberg}}{1989}]{weinberg1989}
{Weinberg} S.,  1989, \mn@doi [Reviews of Modern Physics]
  {10.1103/RevModPhys.61.1}, \href
  {http://adsabs.harvard.edu/abs/1989RvMP...61....1W} {61, 1}

\bibitem[\protect\citeauthoryear{{Winther}, {Koyama}, {Manera}, {Wright}  \&
  {Zhao}}{{Winther} et~al.}{2017}]{winther2017}
{Winther} H.~A.,  {Koyama} K.,  {Manera} M.,  {Wright} B.~S.,   {Zhao} G.-B.,
  2017, \mn@doi [\jcap] {10.1088/1475-7516/2017/08/006}, \href
  {http://adsabs.harvard.edu/abs/2017JCAP...08..006W} {8, 006}

\bibitem[\protect\citeauthoryear{{Wright}, {Koyama}, {Winther}  \&
  {Zhao}}{{Wright} et~al.}{2019}]{wright2019}
{Wright} B.~S.,  {Koyama} K.,  {Winther} H.~A.,   {Zhao} G.-B.,  2019, arXiv
  e-prints, \href {http://adsabs.harvard.edu/abs/2019arXiv190210692W} {}

\bibitem[\protect\citeauthoryear{{Zel'dovich}}{{Zel'dovich}}{1970}]{zeldovich1970}
{Zel'dovich} Y.~B.,  1970, \aap, \href
  {http://adsabs.harvard.edu/abs/1970A26A.....5...84Z} {5, 84}

\makeatother
\end{thebibliography}


\bsp	
\label{lastpage}
\end{document}